\documentclass[11pt,a4paper]{article}
\pdfoutput=1

\usepackage{jcappub}
\usepackage{color}
\usepackage{slashed}
\usepackage{mathtools}
\usepackage{bbm}
\usepackage{graphicx}
\usepackage{epsfig}
\usepackage{subfigure}
\usepackage{bbm}

\newcommand{\be}{\begin{equation}}
\newcommand{\ee}{\end{equation}}

\def\nc{\newcommand}

\def\ie{{i.e.~}}

\def\sfrac#1#2{{\textstyle\frac{#1}{#2}}}
 
\nc{\lsim}{\;\raisebox{-.3em}{$\stackrel{\displaystyle <}{\sim}$}\;}
\nc{\gsim}{\;\raisebox{-.3em}{$\stackrel{\displaystyle >}{\sim}$}\;}

\title{Baryogenesis in the two doublet and inert singlet extension of the Standard Model}

\author[a]{Tommi Alanne} 
\author[b,d]{Kimmo Kainulainen} 
\author[c,d]{Kimmo Tuominen} 
\author[b,d]{Ville Vaskonen} 

\affiliation[a]{{CP}$^{ \bf 3}${-Origins}, University of Southern Denmark,\\
               Campusvej 55, DK-5230 Odense M, Denmark.}
\affiliation[b]{Department of Physics, P.O.Box 35 (YFL), \\ 
                FI-40014 University of Jyv\"askyl\"a, Finland}
\affiliation[c]{Department of Physics, P.O.~Box 64,\\ 
  	        FI-00014 University of Helsinki, Finland.}
\affiliation[d]{Helsinki Institute of Physics, P.O.~Box 64,\\ 
  	        FI-00014 University of Helsinki, Finland.}
\emailAdd{alanne@cp3.sdu.dk}
\emailAdd{kimmo.kainulainen@jyu.fi}
\emailAdd{ville.vaskonen@jyu.fi}
\emailAdd{kimmo.i.tuominen@helsinki.fi}

\abstract{We investigate an extension of the Standard Model containing two Higgs doublets and a singlet scalar field (2HDSM). We show that the model can have a strongly first-order phase transition and give rise to the observed baryon asymmetry of the Universe, consistent with all experimental constraints. In particular, the constraints from the electron and neutron electric dipole moments are less constraining here than in pure two-Higgs-doublet model (2HDM). The two-step, first-order transition in 2HDSM, induced by the singlet field, may lead to strong supercooling and low nucleation temperatures in comparison with the critical temperature, $T_n \ll T_c$, which can significantly alter the usual phase-transition pattern in 2HD models with $T_n \approx T_c$. Furthermore, the singlet field can be the dark matter particle. However, in models with a strong first-order transition its abundance is typically but a thousandth of the observed dark matter abundance.}


\subheader{\small{Preprint: HIP-2016-22/TH, CP3-Origins-2016-031 DNRF90}}

\begin{document}
\maketitle

%
\section{Introduction}
\label{sec:intro}
%

The matter-antimatter asymmetry in the universe presents one of the major quests for particle cosmology. Due to cosmic inflation, such asymmetry cannot be an initial condition for the thermal history of the universe, but calls for a dynamical explanation. The Standard Model (SM) of elementary particle interactions fails in providing a successful mechanism for baryogenesis, and one must look at different extensions of the SM. In this paper we address these issues in the context of a 2HDSM featuring an extended scalar sector with two gauged Higgs doublets and an extra singlet.

Generation of the matter-antimatter asymmetry in connection with the electroweak phase transition, i.e. electroweak baryogenesis, is a particularly appealing scenario due to the possibility of connecting it with the collider experiments. Generic 2HDMs have been studied earlier in connection with the electroweak baryogenesis problem~\cite{Turok:1990zg, Turok:1991uc, Funakubo:1993jg, Davies:1994id, Cline:1995dg, Laine:2000rm, Fromme:2006cm, Cline:2011mm, Dorsch:2013wja}. They provide both a new source of CP violation arising from complex parameters in the 2HDM potential and a strong first-order phase transition arising from the one-loop effective potential. However, observational constraints are placing stringent limits also on 2HDMs~\cite{Cline:2011mm}. Here we show that these constraints are alleviated when the model is further extended by a real scalar singlet field.

A generic feature of 2HDM, also inherited by the 2HDSM, is the danger of generating large flavour changing neutral currents. To avoid these, one has to constrain the Higgs-fermion couplings in one way or the other. Here we choose to work in the context of universal Yukawa alignment, which may be argued for by a requirement that the whole Lagrangian is invariant under the group GL(2,$\mathbbm{C}$) of linear reparametrization transformations in the doublet space. We also use the reparametrization invariance to develop an elegant way explore the vacuum stability and the phase-transition pattern in the model.

In the 2HDM context large CP violation requires that scalar couplings have large complex phases and strong transition requires that couplings are large in magnitude. When combined, these requirements tend to give too large electron and neutron electric dipole moments (EDMs). We will show that the presence of the additional scalar allows for a strong two-step electroweak phase transition, which does not rely on large radiative corrections to the effective potential. This alleviates the burden on the scalar self-couplings and significantly increases the phase space consistent with EDM constraints in the 2HDSM.

The singlet scalar can also be a dark matter (DM) candidate when a discrete $Z_2$ symmetry is imposed to stablize it. However, we will find that a strong first-order phase transition is not consistent with a dominant singlet scalar DM particle. The problem is that a strong two-step transition requires a large coupling between the singlet and doublet sectors and this implies so large annihilation rate for the DM that its relic abundance becomes too small to account for the full observed DM density. This conclusion is generic for all models of this type. 

We observe that two-step transitions may also give rise to {\em too strong} transitions. It is possible that fields get trapped in the metastable minimum so that electroweak symmetry remains unbroken. Also, the latent heat released in the transition may be so large that the transition walls  necessarily become supersonic. However, we find also parameters for which walls may be subsonic, consistent with the electroweak baryogenesis scenario. Overall, we are able to find models that satisfy all observational and experimental constraints and can also give rise to a successful electroweak baryogenesis, accompanied by a subleading DM in the 2HDSM context. 

The structure of the paper is as follows: In Sec.~\ref{sec:model} we introduce the model and discuss the most general GL(2,$\mathbbm{C}$)-reparametrization invariant 2HDSM Lagrangian including Yukawa couplings. Here we also develop methods to study the vacuum stability and the phase-transition patterns in the theory. In Sec.~\ref{sec:results} we first go through the experimental constraints on the model and evaluate the DM relic abundance and the DM search limits on model parameters. We then evaluate the strength of the transition and compute the baryon asymmetry created in the electroweak phase transition. The section is concluded by a study of bubble nucleation in the 2HDSM and in the singlet extension of the SM. In Sec.~\ref{sec:conclusions} we conclude and outline some directions for future research.

%
\section{The model} 
\label{sec:model}
%

We start from the most general two-Higgs-doublet and inert-singlet extension of the SM with the scalar field Lagrangian:
\begin{equation}
\mathcal{L}_{\rm scalar} = \mathrm{Z}^{ij} (D^{\mu}H_i)^\dagger D_{\mu}H_j + \frac{1}{2}(\partial_{\mu}S)^2 - V(H_1,H_2,S)\,,
\label{eq:lag}
\end{equation}
where Z${}^{ij}$ is an arbitrary Hermitian $2\times 2$ matrix and the most general potential is given by
\begin{equation}
\begin{split}
 V(H_1,H_2,S) 
    =& - m_1^2|H_1|^2 - m_2^2|H_2|^2 - \left(m_{12}^2 H_2^{\dagger}H_1 +\; \text{h.c.}\right)
       - \sfrac{1}{2}m_S^2S^2
     \\
 	&+ \lambda_1 |H_1|^4 + \lambda_2|H_2|^4  + \lambda_3 |H_1|^2|H_2|^2
	  + \lambda_4 (H_1^{\dagger}H_2)(H_2^{\dagger}H_1)  \phantom{\sfrac{1}{4}}
     \\
 	&+ \left( \lambda_5(H_2^{\dagger}H_1)^2 
     + \lambda_6  |H_1|^2(H_2^{\dagger}H_1) 
	 + \lambda_7  |H_2|^2(H_2^{\dagger}H_1) \; + \; \text{h.c.} \right)  
     \\
     &+ \sfrac{1}{4}\lambda_SS^4 
     + \sfrac{1}{2}\lambda_{S1}S^2|H_1|^2
	 + \sfrac{1}{2}\lambda_{S2}S^2|H_2|^2
     + \left(\sfrac{1}{2}\lambda_{S12}S^2H_2^{\dagger}H_1+\; \text{h.c.}\right). \;\;
\label{eq:V}
\end{split}
\end{equation}
Both doublets $H_i$ are assumed to be gauged under $\mathrm{SU}(2)_{\mathrm{L}}\times \mathrm{U}(1)_Y$, while the scalar $S$ is a singlet under all SM gauge interactions. The singlet $S$ is a crucial ingredient in the model because it will disentangle the source of a strongly first-order transition from that of sufficiently strong CP violation. 

The Lagrangian (\ref{eq:lag}) is invariant under a reparametrization transformation $\Phi \rightarrow \Phi' \equiv P\Phi$ (and a simultaneous rescaling of $S$), where $P$ is an element of the general linear group GL(2,$\mathbbm{C}$), and $\Phi$ is the Higgs hyperdoublet:
\begin{equation}
 \Phi \equiv (H_1,H_2)^T \,.
\end{equation}
GL(2,$\mathbbm{C}$) is the semidirect product of special linear transformations SL(2,$\mathbbm{C}$) and multiplicative group of dilatations $\mathbbm{C}^\times$. We can always use the dilatation and a hyperbolic SL(2,$\mathbbm{C}$) transformation to bring the kinetic term into the canonical form, Z${}^{ij} \rightarrow {\rm diag}(1,1)$, \ie
$$
\mathrm{Z}^{ij} (D^{\mu}H_i)^\dagger D_{\mu}H_j \quad \rightarrow \quad |D_{\mu}H_1|^2 + |D_{\mu}H_2|^2\,.
$$
The resulting Lagrangian is still invariant under elliptic SL(2,$\mathbbm{C}$) transformations, \ie the usual SU(2) rotations of the doublets.

A generic 2HDM gives rise to unacceptably large flavour-changing neutral currents (FCNCs) and the presence of a singlet does not change the situation. One way to avoid FCNCs is the Yukawa alignment~\cite{Pich:2009sp}, which assumes that both doublets couple to fermions with the same matrix structure (since $S$ is a singlet under SM gauge interactions its couplings to charged SM fermions are excluded):
\begin{equation}
\mathcal{L}_{\rm Yukawa} = y_u C_u^i \bar Q_{\mathrm{L}} \tilde H_i u_{\mathrm{R}} + y_d C_d^i \bar Q_{\mathrm{L}}  H_i d_{\mathrm{R}} 
+ y_\ell C_\ell^i \bar L_{\mathrm{L}} H_i e_{\mathrm{R}} + {\textrm{h.c}},
\end{equation}
where $\tilde H_2 \equiv i\sigma_2 H_2^*$. Here $y_a$ are flavour matrices independent of the doublet index, and $C^a_i$ are doublet-index dependent complex numbers.  In general the alignment may be different in different fermion sectors: $C^a_i \neq C^b_i$. However, for simplicity, we choose to work in the special case of {\em universal Yukawa alignment}, where $C^a_i \equiv C_i$. In this case we can, without a further loss of generality, choose the basis where only the $H_2$ field couples to fermions. This corresponds to setting $C_1 = 0$ and $C_2 = 1$\footnote{This actually involves a rotation and a redefining of the scale of $y^a$ matrices.}, so that: 
\begin{equation}
\mathcal{L}_{\rm Yukawa} = y_u \bar Q_{\mathrm{L}} \tilde H_2 u_{\mathrm{R}} + y_d \bar Q_{\mathrm{L}}  H_2 d_{\mathrm{R}} 
+ y_\ell \bar L_{\mathrm{L}} H_2 e_{\mathrm{R}} + {\textrm{h.c}} \,.
\label{eq:Type-I}
\end{equation}
The choice of basis leading to (\ref{eq:Type-I}) can be effected by an SU(2) rotation of $\Phi$, and it exhausts our remaining freedom to perform elliptic SL(2,$\mathbbm{C}$)-reparametrization transformations after diagonalizing the kinetic term\footnote{Note that in the case of general Yukawa alignment, where $C^a_i \neq C^b_i$, one could still use the SU(2) rotation to set $C^u_1 = 0$, so that up-type quarks couple only to $H_2$. Most of our subsequent analysis would hold also for this scenario, because it is mostly sensitive only to the large top-quark coupling. The only exception is the electron EDM, for which our analysis covers only a part of the full phase space available in the context of general alignment.}. 

Let us stress that while the Yukawa sector (\ref{eq:Type-I}) appears to be of type-I 2HDM, we did not impose any discrete symmetry to derive it. This is why we have kept the $\lambda_6$ and $\lambda_7$ terms in the scalar potential. Note that renormalization does not change the form of the theory; while it both re-introduces a kinetic mixing between doublets and a coupling of $H_1$ to fermions, these changes can be countered by another GL(2,$\mathbbm{C}$) transformation. Also, we point out that the universal Yukawa alignment can be argued for based on reparametrization invariance: only in the context of universal alignment is the {\em complete} Lagrangian including the 2HDM {\em and} Yukawa sectors invariant under GL(2,$\mathbbm{C}$) transformations.

Interestingly, the universal alignment structure arises as a low-energy effective theory in models of dynamical electroweak symmetry breaking~\cite{Simmons:1988fu,Kagan:1991gh,Antola:2009wq,Alanne:2013dra,Geller:2013dla,Hashimoto:2009ty,Fukano:2012qx,Fukano:2013kia}. In the bosonic technicolor~\cite{Simmons:1988fu,Kagan:1991gh,Antola:2009wq}, the ultraviolet theory contains a new gauge theory responsible for dynamically breaking the electroweak symmetry and an elementary scalar doublet $H$, which  communicates the symmetry breaking to the SM fields through its renormalizable Yukawa couplings. At low energies the strong technicolor dynamics is described in terms of an effective Lagrangian for a composite Higgs doublet, which couples with the elementary one through a Lagrangian of the form (\ref{eq:lag}), including the non-trivial kinetic mixing. Only the elementary scalar couples to SM fermions, which naturally introduces Yukawa alignment. Moreover, when the kinetic mixing is removed by a non-unitary transformation, the Yukawa Lagrangian becomes naturally of the {\em universally} Yukawa-aligned form with $C^{a}_i = C_i$.  After a final SU(2) rotation, the model has a diagonal kinetic term, type-I Yukawa sector (\ref{eq:Type-I}), and the most general potential of Eq.~(\ref{eq:V}).

%
\subsection{Reparametrization invariance and tree-level vacuum stability} 
\label{sec:reparametrization}
%

The original Lagrangian with the most general potential, kinetic term, and the universally aligned Yukawa sector has 27 real parameters (not counting the parameters entering the Yukawa-flavour-mixing matrices). We removed the four arbitrary parameters from kinetic terms and three from the complex Yukawa coefficients $C_i$ by the use of the GL(2,$\mathbbm{C}$) invariance of the theory. This still leaves us with 15 real couplings and five real mass parameters in the model potential $V(H_1,H_2,S)$. Our next task is to find out which sets of these parameters correspond to physically viable models with a stable potential. 

We can use the reparametrization invariance to our advantage in constructing the stable potentials. To this end, it is convenient to rephrase the invariance in terms of Lorentz invariance of the potential, written in terms of bilinears formed from hyperdoublets. Following the analysis of Ref.~\cite{Ivanov:2006yq,Ivanov:2007de,Ivanov:2008er}, we define
\begin{equation}
r^\mu \equiv \Phi^\dagger \sigma^\mu \Phi \,,\quad \mbox{where} \quad
\sigma^\mu = (1, \sigma_i) \,.
\end{equation}
The bilinear four-vector $r^\mu$ is positive definite\footnote{Clearly $r_0 = \sum_i |H_i|^2 \ge 0$. Also $r_\mu r^\mu = 4(|H_1|^2|H_2|^2 - (H_1^\dagger H_2)(H_2^\dagger H_1)) \ge 0$\,, by Schwartz inequality.}. That is, $r^\mu$ vectors span the future light cone, LC${}^+$, of a Minkowski space. Thus, in bilinear representation the elliptic and hyperbolic SL(2,$\mathbbm{C}$) basis transformations of fields $\Phi \rightarrow \Phi' \equiv P\Phi$ correspond to proper orthochronous Lorentz transformations $r^\muÊ\rightarrow r'^\mu = {(\Lambda_{\scriptscriptstyle P})^\mu}_\nu r^\nu$, where ${(\Lambda_{\scriptscriptstyle P})^\mu}_\nu\in {\rm SO}(1,3)^+$. In this notation one can rewrite the Higgs potential \eqref{eq:V} in a very compact form:
\begin{equation}
V = - \sfrac{1}{2} m_S^2S^2 - \sfrac{1}{2}M^2_\mu r^\mu 
    + \sfrac{1}{4} r^\mu \lambda_{\mu\nu} r^\nu  
    + \sfrac{1}{4} \lambda_{S\mu} r^\mu S^2 
    + \sfrac{1}{4} \lambda_S S^4\,,
\label{eq:compactPotential}
\end{equation}
where we defined mass and coupling four-vectors
\begin{eqnarray}
M^2_\mu &\equiv&  \left(m_1^2+m_2^2,\, 2 m_{12R}^2,\,
         -2m_{12I}^2,\, m_1^2-m_2^2\right) \,,
         \nonumber \\
\lambda_{S\mu} &\equiv&  \left(\lambda_{S1}+\lambda_{S2},\, 2\lambda_{S12R},\,
         -2\lambda_{S12I},\, \lambda_{S1}-\lambda_{S2}\right)
\end{eqnarray}
and a symmetric coupling tensor
\begin{equation}
\lambda_{\mu\nu} \equiv 
\left( \begin{array}{cccc}
    \lambda_1 + \lambda_2 + \lambda_3\quad  &   \lambda_{6R} + \lambda_{7R}\quad  & 
    -\lambda_{6I} + \lambda_{7I}\quad        &   \lambda_1 - \lambda_2          \\ [1mm]
     \lambda_{6R} + \lambda_{7R}        &   \lambda_4 + 2 \lambda_{5R}   & 
    -2\lambda_{5I}                     &   \lambda_{6R} - \lambda_{7R}     \\ [1mm]
    -\lambda_{6I} + \lambda_{7I}        &  - 2\lambda_{5I}               & 
     \lambda_4 -2 \lambda_{5R}         &  -\lambda_{6I} - \lambda_{7I}     \\[1mm]
    \lambda_1-\lambda_2                &   \lambda_{6R} - \lambda_{7R}   &
   -\lambda_{6I} - \lambda_{7I}         &   \lambda_1 + \lambda_2  - \lambda_3
\end{array} \right)\, ,
\label{lambda}
\end{equation}
where the subscripts $R$ and $I$ refer to the real and imaginary parts of the couplings, respectively. The reparametrization invariance is manifest in Eq.~\eqref{eq:compactPotential}, because $V$ depends only on Lorentz-invariant products of vectors and tensors. This form is particularly suitable for a study of the vacuum stability and phase-transition patterns of the model.

First consider the direction $S=0$ in the potential~\eqref{eq:compactPotential}. Here the term $r^\mu \lambda_{\mu\nu} r^\nu$ must be bounded from below. In~\cite{Ivanov:2006yq} it was shown that this is the case precisely when $\lambda_{\mu\nu}$ is positive definite in the future light cone. That is, all stable potentials can be written as $\lambda_{\mu\nu}Ê\equiv {\Lambda_\mu}^\alpha \lambda^{\scriptscriptstyle D}_{\alpha\beta} {\Lambda^\beta}_\nu$, where ${\Lambda_\mu}^\alpha$ is an SO(1,3)$^+$ transformation and
\begin{equation}
\lambda^{\scriptscriptstyle D}_{\alpha\beta} = {\rm diag} (\lambda^{\scriptscriptstyle D}_{00},-\lambda^{\scriptscriptstyle D}_{11},-\lambda^{\scriptscriptstyle D}_{22},-\lambda^{\scriptscriptstyle D}_{33}),\quad {\rm with} \quad \lambda^{\scriptscriptstyle D}_{00}Ê> 0 \quad {\rm and} \quad  \lambda^{\scriptscriptstyle D}_{00} >  \lambda^{\scriptscriptstyle D}_{ii} \,.
\label{eq:positivityconditions2}
\end{equation}
Note that the four parameters in $\lambda^{\scriptscriptstyle D}_{\alpha\beta}$ together with the six parameters in ${\Lambda^\beta}_\nu$ add up to the ten real degrees of freedom in the most general 2HDM potential. Second, if we set $r^\mu = 0$ ($H_1 = H_2 =0$), we see that we must have
\begin{equation}
 \lambda_S > 0 \,.
\end{equation}
Finally, we have to consider the directions where both $S$ and $H_i$ are nonzero. First, if the vector $\lambda_{S\mu}$ of couplings which mix $S$ and $H_i$ lies in the future light cone, $\lambda_{S\mu}\in \mathrm{LC}^+$, i.e.
\begin{equation}
\lambda_{S\mu}\lambda_{S}^\mu 
= 4 \left( \lambda_{S1}\lambda_{S2} - |\lambda_{S12}|^2 \right) > 0 
\quad {\rm and} \quad \lambda_{S}^0 = \lambda_{S1} + \lambda_{S2} > 0 \,, 
\label{eq:positivityoflambdaS}
\end{equation}
then the mixing term $\sfrac{1}{4}\lambda_{S\mu}r^\mu S^2$ in the potential, Eq.~\eqref{eq:compactPotential},  is always positive, and no new conditions arise. However, if $\lambda_{S\mu}\notin \mathrm{LC}^+$, there are always directions $r^\mu \in \mathrm{LC}^+$ along which the product $\lambda_{S\mu}r^\mu$ is negative. If we in such cases rewrite the quartic part of the potential as
\begin{equation}
V_{\rm quartic} = 
\frac{1}{4} r^\mu 
\left( \lambda_{\mu\nu} - \frac{1}{4\lambda_S}\lambda_{S\mu}\lambda_{S\nu}\right)r^\nu  
    + \frac{1}{4}\lambda_S (S^2 + \frac{\lambda_{S\mu} r^\mu}{2\lambda_S})^2 \,,
\label{eq:compactPotential2}
\end{equation}
we see that the potential is most negative as a function of $S$ along direction $2\lambda_S S^2 = - \lambda_{S\mu} r^\mu$. In this subspace, the potential reduces to the form 4$V_{\rm quartic} = r^\mu  \lambda^{\scriptscriptstyle S}_{\mu\nu} r^\nu$, with a new coupling matrix:
\begin{equation}
\lambda^{\scriptscriptstyle S}_{\mu\nu} 
\equiv \lambda_{\mu\nu} - \frac{1}{4\lambda_S}\lambda_{S\mu}\lambda_{S\nu} \,. 
\label{eq:lambdaS}
\end{equation}
The matrix $\lambda^{\scriptscriptstyle S}_{\mu\nu}$ is not in general diagonalizable by an SO$(1,3)^+$ rotation, and it may have also complex eigenvalues. Unfortunately we cannot restrict its properties like we did for $\lambda_{\mu\nu}$, because $\lambda^{\scriptscriptstyle S}_{\mu\nu}$ does not need not be positive definite in the entire future light cone, but only in the subset of $\mathrm{LC}^+$ where $\lambda_{S\mu}r^\mu$ is negative. Instead of covering the full range of possibilities, we will require a sufficient (but not necessary) condition that $\lambda^S_{\mu\nu}$ is positive definite in the future light cone whenever $\lambda_{S\mu}\notin \mathrm{LC}^+$. 

We now have the recipe to construct the space of stable potentials: we first choose a $\lambda^{\scriptscriptstyle D}_{\alpha\alpha}$ which satisfies Eqs.~\eqref{eq:positivityconditions2}. Then we generate a vector $\lambda^{\scriptscriptstyle D}_{S\mu}$ and check if it satisfies the positivity constraint \eqref{eq:positivityoflambdaS}, or if the matrix $\lambda^{\scriptscriptstyle S}_{\mu\nu}$ in Eq.~\eqref{eq:lambdaS} is positive definite in the subset of LC${}^+$ where $\lambda_{S\mu}r^\mu < 0$. Having found an acceptable set, we generate a random Lorentz transformation and define 
\begin{equation}
\lambda_{\mu\nu} \equiv 
{\Lambda_\mu}^\alpha \lambda^{\scriptscriptstyle D}_{\alpha\beta}{\Lambda^\beta}_\nu 
\quad {\rm and}Ê\quad
\lambda_S^\mu \equiv {\Lambda^\mu}_\nuÊ(\lambda^{\scriptscriptstyle D}_{S})^\nu\,,
\end{equation}
where ${\Lambda^\mu}_\nu \in {\rm SO}(1,3)^+$.

Let us comment on the role of the kinetic and Yukawa terms in the above construction of the potential. We implicitly assumed that kinetic term becomes diagonal, and the Yukawa term becomes of type-I form in the final frame, after the Lorentz transformation. Thus, they necessarily must be nontrivial in the original, diagonal frame. Indeed, the six degrees of freedom ``missing$"$ in the diagonal potential are in this frame evenly divided between the $C_i$ coefficients in the Yukawa Lagrangian and the mixing parameters in the kinetic term, which in the bilinear notation can be written as
\begin{equation}
K_\mu (D^\alpha \Phi)^\dagger \sigma^\mu (D_\alpha \Phi) \,.
\end{equation}
Here $K_\mu$ is some positive definite, but otherwise arbitrary four-vector of unit length (here $\alpha$ refers to the usual space-time indices and $\mu$ to potential indices)\footnote{In the most general case $K_\mu$ has four free parameters, but one parameter is here assumed to be removed by a dilatation, used to bring the length of $K_\mu$ to unity.}. It is amusing to see that exactly a Lorentz boost (a hyperbolic SL(2,$\mathbbm{C}$) transformation on fields) is needed to bring an arbitrary $K_\mu$ into the canonical form: $K_\mu \rightarrow (1;0,0,0)$, after which the kinetic term is manifestly invariant under Lorentz rotations (elliptic SL(2,$\mathbbm{C}$) transformations on fields). These boosts and rotations into the canonical frame activate the whole ${\rm SO}(1,3)^+$ group discussed above and thus create all physically viable Lagrangians with bounded potentials.

%
\subsection{Spontaneus symmetry breaking} 
\label{sec:spntaneousbreaking}
%

Since we are interested in the cases where the singlet scalar $S$ is a DM candidate, we restrict our considerations to the cases where the potential is unbroken in S direction and hence $Z_2$ symmetric at low temperatures. However, to enhance the strength of the latter phase transition, we also need the $S$ symmetry to be broken at high temperatures before the symmetry breaks in the doublet direction. To this end, we must have negative quadratic term in the $S$ direction in the potential ($m_S^2>0$). This implies that there may be other minima away from the $\langle S\rangle=0$ vacuum, and we must check that none of these minima is the global one at zero temperature. The extremization conditions are:
\begin{equation}
\frac{\partial V}{\partial S}  = S\zeta_S = 0 \,, 
\quad \mbox{where} \quad 
\zeta_S \equiv - m^2_S + \lambda_S S^2 + \sfrac{1}{2}\lambda_{S\mu}r^\mu\,,
\label{eq:statcond1}
\end{equation}
and
\begin{equation}
\frac{\partial V}{\partial H_i^\dagger}  = \sigma_{ij}^\mu H_j \zeta_\mu = 0 \,,
\quad \mbox{where} \quad 
\zeta_\mu = - \sfrac{1}{2}\left(M^2_\mu - \sfrac{1}{2}\lambda_{S\mu}S^2\right) 
            + \sfrac{1}{2}\lambda_{\mu\nu} r^\nu\,.
\label{eq:statcond2}
\end{equation}
Eqs.~\eqref{eq:statcond2} are complex, so we have five equations relating vacuum fields to the parameters of the theory. Our goal is to have an unbroken singlet and a broken neutral doublet vacuum at zero temperature. In~\cite{Ivanov:2006yq} it was shown that the neutral and charged vacua cannot coexist in the pure 2HDM case. When $S = 0$, our potential reduces to the pure 2HDM case, and so the above statement applies here as well. Therefore, if we find a neutral vacuum, we know it is the global one in the doublet space and the charged extremum may at best be a saddle point. Moreover, even with $S \neq 0$, the doublet-symmetry-breaking pattern is formally similar to the pure 2HDM case, only with an effective mass parameter
\begin{equation}
 - m^2_\mu \equiv  -M^2_\mu + \sfrac{1}{2}\lambda_{S\mu}S^2.
\end{equation}
This implies that also any neutral minimum with $S$ breaking is the lowest one in the doublet space.

The most general neutral vacuum in 2HDM field space is given by
\begin{equation}
 \langle H_1 \rangle  = 
 \frac{1}{\sqrt{2}}\left(\begin{array}{cc} 
   0 \\
   v_1 e^{i\theta_1}
  \end{array} \right), 
  \quad {\rm and}Ê\quad
\langle H_2 \rangle  = 
 \frac{1}{\sqrt{2}}\left(\begin{array}{cc} 
   0 \\
   v_2 e^{i\theta_2}
  \end{array} \right) \,.
\label{eq:minimumdoublets}
\end{equation}
Of course the local $\mathrm{SU}(2)$-gauge invariance guarantees that only the relative phase $\theta \equiv \theta_2-\theta_1$ is physical, and one could rotate for example $\theta_1 \rightarrow 0$. Explicitly we find that this corresponds to
\begin{equation}
r_0^\mu = (v_1^2+v_2^2, 2v_1v_2 \cos\theta,2v_1v_2 \sin\theta, v_1^2-v_2^2)\,,
\end{equation}
so that $r_0^2 = 0$ as it should for a neutral vacuum~\cite{Ivanov:2006yq}. This construction actually only ensures that the special point \eqref{eq:minimumdoublets} is an {\em extremum} of the potential. To check that this extremum is a also a minimum, one needs to compute second derivative matrices of the potential corresponding to scalar field masses and require that there are no negative eigenvalues. For the mass of the physical excitation in the singlet direction, this corresponds to requiring that
\begin{equation}
\frac{{\rm d}^2 V}{{\rm d}S^2}\biggr\vert_{\rm{vev}}
=-m_S^2 + \sfrac{1}{2}\lambda_{S\mu}r_0^\mu \; \equiv  \; M_S^2  \; >  \; 0 \,,
\label{eq:DMmasscond}
\end{equation}
where the left hand side is evaluated at the extremum $\langle S\rangle = 0$ and $\langle H_i\rangle\neq 0$ as defined in Eq. \eqref{eq:minimumdoublets}. These requirements on the spectrum allow one to set the Lagrangian mass parameters in terms of physical masses and vacuum expectation values (vevs) of the fields. However, it still remains to check that the vacuum with $S=0$ is the global minimum for any given set of parameters. It is straightforward to show that the value of the potential at the desired $\langle S\rangle=0,\langle H_i\rangle\neq 0$ vacuum is
\begin{equation}
V(\langle S\rangle=0,\langle H_i\rangle \neq 0) = -\frac{1}{4} r_0^\mu \lambda_{\mu\nu}r_0^\nu \,.
\label{eq:DMmasscond2}
\end{equation}
In the direction $H_i=0$, the potential has at least a directional minimum at $\langle S\rangle^2 = m_S^2/2\lambda_S$, and the value of the potential in this directional minimum is
\begin{equation}
V(\langle S\rangle\neq0,\langle H_i\rangle=0) = -\frac{1}{4\lambda_S} (M_S^2 - \sfrac{1}{2}\lambda_{S\mu}r_0^\mu)^2 \,.
\label{eq:DMmasscond3}
\end{equation}
We impose the condition that the minimum in Eq.~\eqref{eq:DMmasscond2} is below that in Eq.~\eqref{eq:DMmasscond3}. Finally, there is an extremum where both $\langle S\rangle \neq 0$ and $\langle H_i\rangle\neq 0$, but this is a local maximum.

In addition to the massive scalar $S$, the physical spectrum contains the usual states arising from the two Higgs doublets: the three neutral scalars $h_0$, $H_0$ and $A_0$, and two charged scalars $H^\pm$. The diagonalization of the mass matrices is presented in detail in Appendix \ref{massappendix}. The lightest neutral scalar state $h_0$ is identified with the 125 GeV Higgs particle observed LHC, while the masses of the heavier neutral and charged scalar states are constrained to lie above the current limits.

%
\subsection{Finite-temperature potential} 
\label{finiteTpot}
%

The final ingredient we need for our analysis is the effective potential at finite temperature. Here we only consider the leading corrections to the potential, which bring about the symmetry restoration at high temperatures. In the high-temperature limit, these corrections are accounted for by the thermal masses:
\begin{equation}
m_a^2(T) =  -m_a^2 + c_a\frac{T^2}{12},
\label{eq:quartic}
\end{equation}
where ($a=1, 2, 12, S$):
\begin{eqnarray}
c_1 &=& c_{\rm SM} + 6 \lambda_1 + 2 \lambda_3 + \lambda_4 + \sfrac{1}{2}\lambda_{S1}
\nonumber\\
c_2 &=& c_{\rm SM} + 6 \lambda_2 + 2 \lambda_3 + \lambda_4 + \sfrac{1}{2}\lambda_{S2}
\nonumber\\
c_{12} &=& c_{\rm SM} + 3 \lambda_{6R} + \lambda_{7R} + \sfrac{1}{2}\lambda_{S12R}
- \mathrm{i} (3 \lambda_{6I} - \lambda_{7I} + \sfrac{1}{2}\lambda_{S12I} )
\nonumber\\
c_S &=& 3 \lambda_S + 2 (\lambda_{S1} + \lambda_{S2}) \,,
\end{eqnarray}
and
\begin{eqnarray}	
c_{\rm SM} = \frac{9}{4}g_L^2 + \frac{3}{4}g_Y^2 + 3y_t^2
\end{eqnarray}
is the common SM contribution from the SU(2)$_{\mathrm{L}}$ and U(1)${}_Y$ gauge fields and the top quark.

%
\section{Results}
\label{sec:results}
%

We now test our model against experimental constraints and for the consistency of its cosmological predictions. We scan the parameter space by solving $m^2_1$, $m_2^2$ and $m^2_{12}$ from the vacuum conditions and create Monte Carlo chains in the coupling constant space by the algorithm described in the previous section. We first subject the models to various theoretical and experimental constraints. For models that pass these constraints, we compute the DM abundance and the strength of the phase transition. We find that the model can provide either the observed DM abundance or a strong transition, but not both simultaneously. For the points providing a strong transition, we compute the prediction for the baryon-to-entropy ratio $\eta_B$. We show that the parameter space where $\eta_B$ is sufficiently large is strongly constrained but not excluded by the current experimental limit on the electric dipole moment of the electron.  Finally, we compute the nucleation temperatures $T_n$ and show that typically, and in particular for strong transitions, $T_n$ is much smaller than the critical temperature $T_c$. 

%
\subsection{Theoretical and experimental constraints} 
\label{sec:experiments}
%

The couplings between scalar fields tend to run strongly, which typically leads to a relatively low cut-off scale for the validity of the effective theory. To be specific, we demand that our model is consistent up to $\Lambda = 1.5$ TeV, i.e. we check that the vacuum is stable and that all couplings remain perturbative\footnote{For perturbativity we use the upper limit $(4\pi)^2$ on all scalar couplings. Over large portion of our data points the couplings will remain small also at larger scales $\mu\gg \Lambda$; the value 1.5 TeV is chosen for concreteness.} up to $\Lambda$. The 1-loop renormalization-group equations needed for this calculation are summarized in Appendix \ref{RGEs}. Moreover, at zero temperature we impose
\begin{equation} 
|\langle H_1\rangle|^2+|\langle H_2\rangle|^2=v_0^2,\quad \langle S\rangle=0
\label{T0vacuum}
\end{equation} 
and check that this choice gives the global minimum of the potential as detailed in previous section.

Second, we address the current experimental constraints. We only accept points that give a light scalar of mass $m_{h_0}=125$ GeV, and for which the heavy scalars satisfy the mass limits from direct searches at LEP on charged particles, $m_{H^+}>500$ GeV~\cite{Heister:2002ev}, and at LHC on heavy neutral scalars, $m_{H_0},m_{A_0}>600$ GeV~\cite{Aad:2012an}. For consistency, we also constrain the heavy scalar masses to be below the cut-off $\Lambda = 1.5$ TeV. We also take into account the electroweak precision data using the $S$ and $T$ parameters~\cite{Peskin:1990zt}. The necessary formulae for computing $S$ and $T$ can be extracted from~\cite{Grimus:2008nb}, where the oblique parameters have been calculated for models with extra doublets and extra singlets.  The current values of $S$ and $T$ parameters are $S=0.00\pm0.08$, $T=0.05\pm 0.07$ with a correlation factor $\rho=0.90$~\cite{Agashe:2014kda}. We accept points that lie within the $2\sigma$ region in the $(S,T)$ plane.

Moreover we take into account constraints on the Higgs boson couplings to SM particles using the signal-strength data from the LHC and Tevatron experiments~\cite{Aaltonen:2013ioz,Khachatryan:2014jba,Aad:2015gba}. First, we impose the constraint from the invisible width of the Higgs boson from LHC at $2\sigma$ level: 
\begin{equation}
R_{\rm I} = \frac{\Gamma_{\rm I}}{\Gamma_{\rm I} + \Gamma_{\rm SM,\,tot}} < 0.020,
\end{equation}
which here implies $\Gamma_{h_0\to SS}\lsim 1.0$ MeV. Second, we allow for modifications to the Higgs couplings and parametrize the deviations from the SM with parameters $a_f$ and $a_V$,
\begin{equation}
\mathcal{L}_{\rm eff} = a_V\left(\frac{2M_W^2}{v_0} h_0 W^+ W^- + \frac{M_Z^2}{v_0}h_0 Z Z \right) - a_f \sum \frac{m_\psi}{v_0} h_0 \overline \psi \psi \;.
\end{equation}
We accept points which are within $2\sigma$ from the best fit values
\begin{equation}
a_V = 0.993\,,\quad a_f = 0.968\,,
\end{equation}
which we obtain by performing a $\chi^2$ fit to the Higgs boson signal strength data. We have neglected the imaginary part of $a_f$ and the $h_0H^+H^-$ coupling in the fit, since we have checked that these couplings are very small\footnote{As a comparison, we have performed the $\chi^2$ fit of $a_f$ and $a_V$  with maximum values of the imaginary part of $a_f$ and $h_0H^+H^-$ coupling 
for out data points, and we find that the best-fit value does not change significantly.}.

%
\subsection{Dark matter abundance and direct detection limits}
\label{sec:DMlimits}
%

We compute the relic abundance of the $S$ bosons for all models passing the experimental constraints  described above. We assume that $S$ is a thermal relic, i.e. we assume that at least some the portal couplings $\lambda_{S1}$, $\lambda_{S2}$ and $\lambda_{S12}$ are sufficiently large (larger than about $10^{-7}$~\cite{Kainulainen:2016vzv}). We then apply the standard freeze-out formalism~\cite{Gondolo:1990dk}, employing the accurate approximation scheme introduced in~\cite{Cline:2013gha}. The relevant annihilation channels of our WIMP candidate are
\begin{equation}
SS\rightarrow h_0h_0,\, H_0H_0,\, A_0A_0,\, h_0H_0,\, h_0A_0,\, H_0A_0,\, H^+H^-,\, W^+W^-,\, ZZ,\, \bar{f}f.
\end{equation}
Cross-sections for all these processes are given in a very compact form in Appendix~\ref{cross-sections}. To treat cases where $S$ is a subdominant DM candidate, we define the ratio~\cite{Cline:2012hg}
\begin{equation} 
f_{\rm{rel}}=\frac{\Omega_Sh^2}{0.12}\,,
\end{equation} 
which expresses how large fraction of the observed DM abundance is in form of $S$ bosons. All annihilation channels are directly proportional to the couplings between the $S$ boson and the Higgs fields, although the precise dependence is rather complicated (see Appendix~\ref{cross-sections}). Since the relic abundance is roughly proportional to the inverse of the annihilation cross section, large (small) couplings corresponds to a small (large) relic abundance. 

Direct search limits for $S$ follow from the bound on the spin-independent cross section for $S$ scattering off nucleons. It is given by
\begin{equation}
\label{eq:sigmaSI}
\sigma_{\mathrm{SI}}=\frac{\lambda_{\rm eff}^2f_N^2}{4\pi}\frac{\mu m_N^2v_0^2}{m_h^4 M_S^2},
\end{equation}
where $m_N=0.939$~GeV is the nucleon mass, $\mu=m_N M_S/(m_N+M_S)$ is the reduced mass of the nucleon--scalar system, and $f_N \approx 0.30$~\cite{Cline:2013gha} gives the strength of the Higgs--nucleon coupling: $g_{h_0N\bar N} = f_N m_N/v_0$. Finally, the effective $SSh_0$ coupling is given by
\begin{eqnarray}
\lambda_{\rm eff} &\equiv&\frac{1}{2}\Big[ 
\big( R_{N_{44}} \cos\beta \sin\theta  - R_{N_{42}} \cos\beta \cos\theta 
                                       + R_{N_{41}} \sin\beta \big)\lambda_{S12I} 
\nonumberÊ\\
&& \,\;+
\big( R_{N_{42}} \cos\beta \sin\theta - R_{N_{44}} \cos\beta \cos\theta
                                      - R_{N_{43}} \sin\beta \big)\lambda_{S12R} 
\nonumberÊ\\
&& \,\;
+ \cos \beta \big( R_{N_{43}} \cos\theta + R_{N_{41}} \sin\theta \big) \lambda_{S1} 
- \sin \beta R_{N_{44}} \lambda_{S2} \Big]\,,
\end{eqnarray}
where $\beta$ is the vacuum mixing angle and $\theta$ the phase between the doublet vevs, and $R_{N_{ij}}$ are the components of the  $4\times 4$-mixing matrix between the neutral scalar fields given in Eq.~(\ref{eq:neutraldiagonalizingmatrix}).

\begin{figure}
\begin{center}
\includegraphics[width=0.7\textwidth]{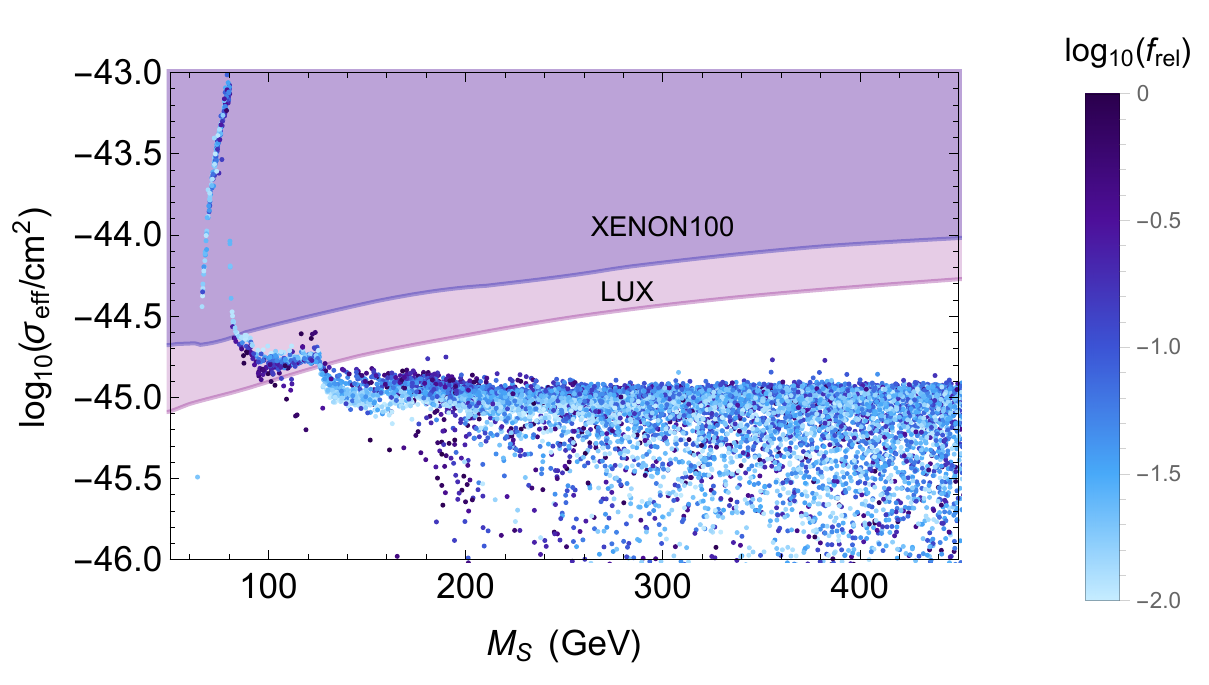} \hspace{0.1cm}
\caption{Shown is the LUX bound on a set of models passing all experimental constraints described in Sec.~\ref{sec:experiments} and for which 
$f_{\mathrm{rel}}\geq 0.01$.} 
\label{fig:LUX}
\end{center}
\end{figure}

Currently the most stringent limit for $\sigma_{\mathrm{SI}}$ come from the LUX experiment \cite{Akerib:2015rjg}. However, in the case of a subdominant DM, the LUX bound is not directly related to $\sigma_{\mathrm{SI}}$ shown in Eq.~\eqref{eq:sigmaSI}. Instead, assuming that all DM components cluster similarly, the actual signal strength from $S$ bosons is suppressed by the fraction $f_{\rm rel}$. The relevant quantity to compare with the LUX limit then is 
\begin{equation}
\label{eq:sigmaeff}
\sigma_{\mathrm{eff}}=f_{\mathrm{rel}}\sigma_{\mathrm{SI}}.
\end{equation}
We show the effect of this limit on the models passing all experimental bounds on Fig.~\ref{fig:LUX}. Note that the scatter in $\sigma_{\rm eff}$ as a function of $m_S$ is much larger here than in the singlet extension of the SM~\cite{Cline:2012hg}. Note how also the models with relatively low abundance are constrained by the LUX data.

%
\subsection{Electroweak phase transition}
\label{sec:EWPT}
%

One of the necessary conditions for a successful baryogenesis scenario is the departure from thermal equilibrium. In EWBG this is provided by a strong first-order phase transition. In pure 2HDMs, with only leading thermal corrections (i.e. with thermal masses), the phase transition is of second order. While full one-loop effects may induce a first-order transition, it still tends to be rather weak~\cite{Cline:2011mm,Dorsch:2013wja}. In a model with a singlet scalar, a strong transition may take place with just the leading corrections~(\ref{eq:quartic}), given a specific two-step symmetry-breaking pattern~\cite{Espinosa:2011ax}. This means that, when passing from high towards low temperatures, a minimum of the potential is generated first along the singlet direction. At lower temperature, then, the potential develops a minimum where the doublets have non-zero vevs while the singlet symmetry is restored, $(\langle h_1\rangle,\langle h_2\rangle; \langle s\rangle)=(0,0;0)\rightarrow (0,0;w(T))\rightarrow (v_1(T),v_2(T);0)$,  ensuing the actual electroweak phase transition. This pattern of minima should of course develop in such a way that the true ground state at $T=0$ is given by Eq.~\eqref{T0vacuum}. 

Including the thermal mass corrections of Eq.~(\ref{eq:quartic}) to the potential, we find the transition temperatures.
In particular the electroweak transition temperature, $T_c$, where the two minima
are degenerate is determined by
\begin{equation}
V(0,0,w_c,T_{c}) = V(v_{1c},v_{2c},0,T_c),
\label{eq:criticalT}
\end{equation}
where $w_{c}=w(T_{c})$ and $v_{ic}=v_{i}(T_{c})$. To determine when the transition is strong enough, we require, as usual, that
\begin{equation}
   v_c/T_c \ge 1 \,,
  \label{eq:stongEWPT}
\end{equation}
where $v_{c}=\sqrt{v_{1c}^{2}+v_{2c}^{2}}$. It turns out not to be possible to have simultaneously strong transition and dominant DM candidate in the model with $f_{\rm rel}\approx1$. This is because having a strong transition requires that at least some of the mixing couplings $\lambda_{Si}$ are large. However, as we noted above, large mixing introduces large annihilation cross sections for $S$ bosons and hence small relic abundances. Qualitatively the behaviour is the same as in the pure singlet model~\cite{Cline:2012hg,Cline:2013gha}: a strong first-order EWPT and large $f_{\rm{rel}}\approx 1$ are realized in different portions of the parameter space. 

\begin{figure}
\begin{center}
\includegraphics[width=0.475\textwidth]{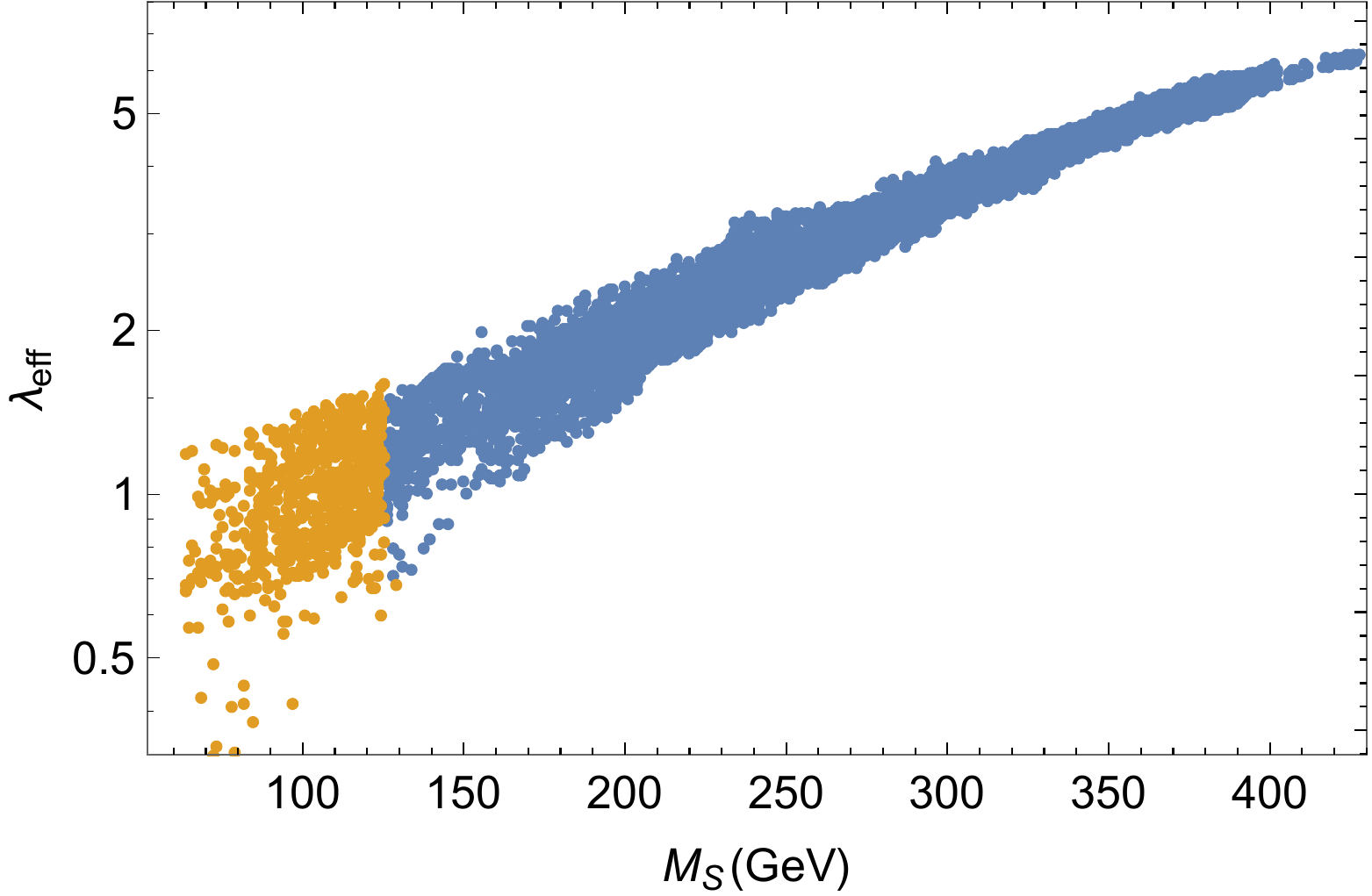} \hspace{1mm}
\includegraphics[width=0.485\textwidth]{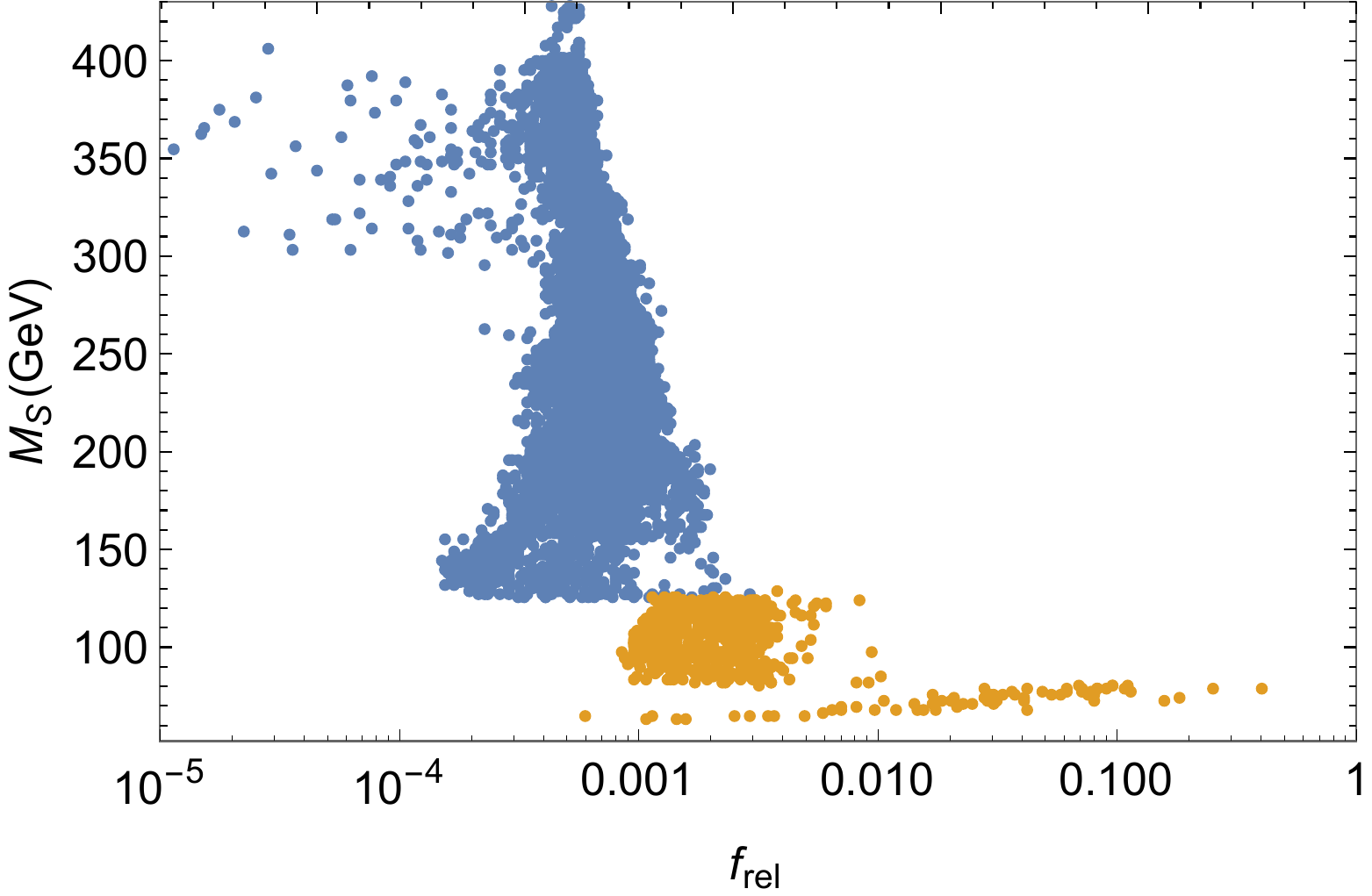}
\caption{Scanned data points which give a strong first-order EWPT. Yellow points are excluded by direct DM searches.}
\label{EWPT}
\end{center}
\end{figure}

Left panel of Fig.~\ref{EWPT} shows the correlation between the DM particle mass $M_S$ and the effective coupling $\lambda_{\rm{eff}}$ for models with strong EWPT. Accepted values of $\lambda_{\rm eff}$ increase as a function of $M_{S}$ because both quantities are linearly proportional to the mixing couplings $\lambda_{Si}$. Right panel of Fig.~\ref{EWPT} shows the correlation between $M_{S}$ and the relative relic abundance $f_{{\rm rel}}$ for the same set. Note that for $M_{S}$ between the Higgs resonance and $M_{W}$, both a strong transition and a relatively large DM abundance, $f_{\rm rel}\approx 0.15$ could be obtained, but this region is now excluded by LUX. 

The fact that LUX most strongly constrains models with {\em small} $\lambda_{\rm eff}$ is because $f_{\rm rel}$ is roughly inversely proportional to the square of $\lambda_{\rm eff}$, and this to large extent cancels the direct $\lambda_{\rm eff}$ dependence in $\sigma_{\mathrm{SI}}$. Larger mass region beyond the Higgs resonance is less constrained because the DM-number density falls with increasing mass.

%
\subsection{Electron EDM constraint}
\label{sec:EDM}
%

\begin{figure}
\begin{center}
\includegraphics[width=0.47\textwidth]{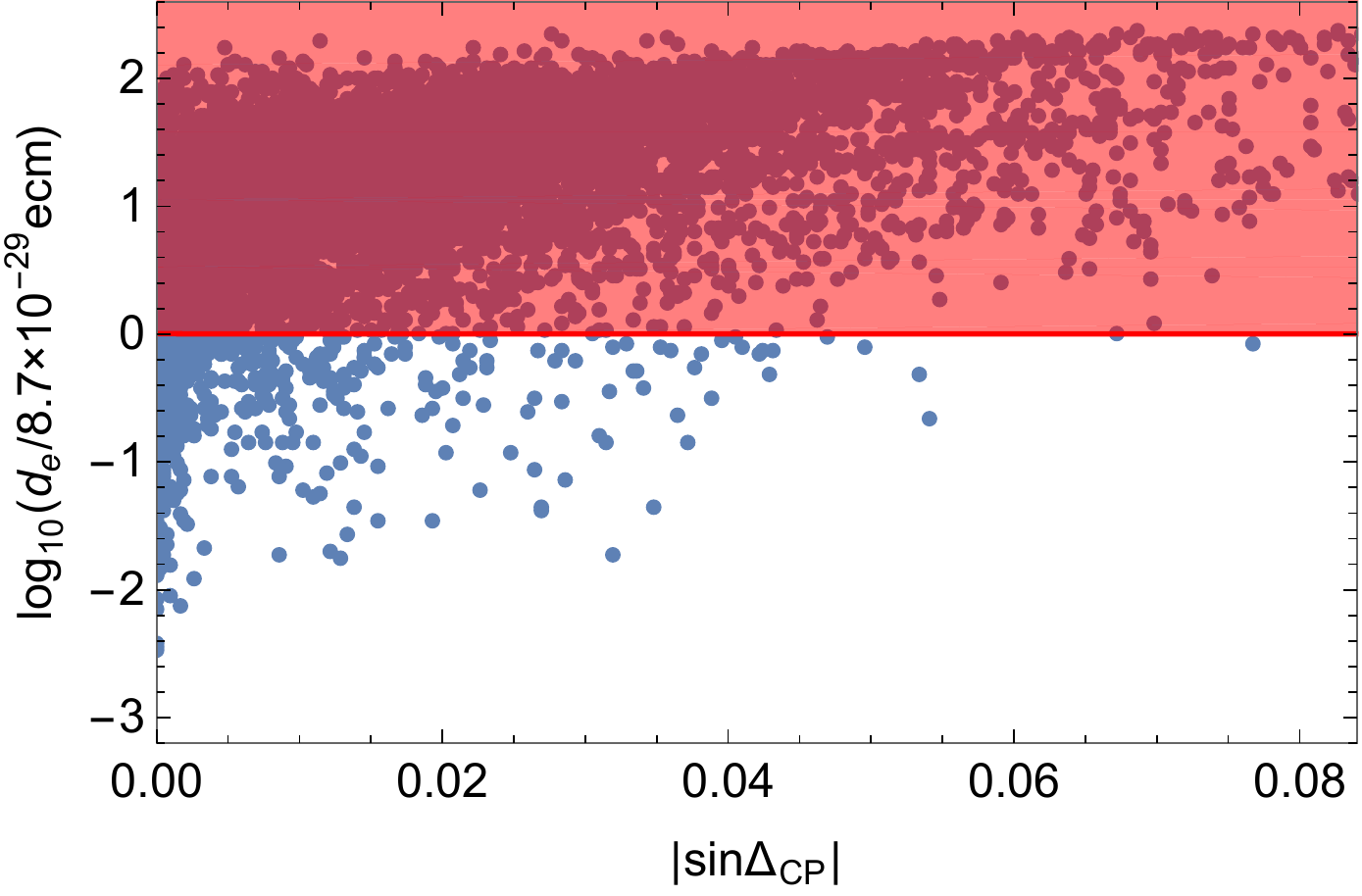}
\caption{Scatter plot of all models with strong enough EWPT as a function of the mixing parameter $\sin\Delta_{\rm CP}$ and the electron EDM $d_e$. The red region is excluded by the eEDM limit. }
\label{eEDM}
\end{center}
\end{figure}

The non-observation of electric dipole moments (EDMs) of electrons, neutrons and atoms gives stringent bounds on CP-violating interactions in multi-Higgs models. As shown in~\cite{Inoue:2014nva}, currently the most stringent bound for 2HDMs arises from the electron EDM, for which the ACME experiment gives an upper limit 
\begin{equation} 
|d_e|< 8.7\times10^{-29}e{\rm cm} \,,
\end{equation}
with $90\%$ confidence level~\cite{Baron:2013eja}. We calculate $d_e$ for the points which give a strong first-order EWPT using the results from Ref.~\cite{Abe:2013qla}, where Barr--Zee type contributions to fermionic EDM were calculated in 2HDM. These results are directly applicable here as well, because the singlet scalar $S$  does not directly couple to gauge fields. In Fig.~\ref{eEDM}, we show the  distribution of models passing all previous cuts as a function of $d_e$ and the neutral scalar mixing matrix element $R_{N_{42}}$, which expresses the projection of $h_0$ to complex part of the second doublet:
\def\sdcp{\sin \Delta_{\rm CP}}
\begin{equation}
R_{N_{42}}\equiv \langle H_{2I}^0| h_0\rangle \equiv \sdcp \,. 
\label{eq:RN42}
\end{equation} 
$\sin\Delta_{\rm CP}$ is given in terms of the various mixing angles in Eq.~(\ref{eq:neutraldiagonalizingmatrix}). The red region is excluded by the electron EDM constraint. Small $d_e$ naturally correlates with small $\sin\Delta_{\rm CP}$, because the size of $\sin\Delta_{\rm CP}$ is proportional to the size of the CP-violating mixing in the model.

%

%
\subsection{Baryogenesis}
%

The actual baryogenesis mechanism in our model relies on CP-violating interactions of the top quark with the expanding phase-transition walls. The CP violation comes directly from the spatial evolution of the complex phases of the Higgs field $H_2$, which renders the top mass a complex-valued function of the spatial coordinate across the wall. The first step for us is then to work out the evolution of the scalar fields over the bubble wall.

We shall approximate the true phase-transition-wall profile in the usual way, by the stationary path that extremizes the Euclidean one-dimensional action 
\begin{equation} 
\int {\rm d}z \left( |D_z H_1| + |D_z H_2| + \frac{1}{2}|\partial_z S| + V + \dots \right),
\end{equation} 
between two degenerate minima at critical temperature $T=T_c$, for which the condition (\ref{eq:criticalT}) holds. The covariant derivatives involve the classical $Z_\mu$ field: $D_\mu = \partial_\mu - ig/(2\cos\theta_{\rm W})Z_\mu$. We write the neutral components of the doublets as $h_j e^{i\varphi_j}$ and observe that the effective potential can depend only on the relative phase $\varphi \equiv \varphi_1-\varphi_2$. Following Ref.~\cite{Cline:2011mm}, we work in the gauge $Z_\mu = 0$, whereby we need to account for four fields: $h_1,h_2,S$ and $\varphi$, while solving the path. The relevant reduced action is 
\begin{equation}
S_1 = \int {\rm d}z \left( \sum_i \frac{1}{2}(\partial_z h_i)^2 
+ \frac{1}{2}(\partial_z S)^2
 + \frac{1}{2}\frac{h_1^2h_2^2}{h_1^2+ h_2^2} (\partial_z \varphi)^2 
 + V(h_1,h_2,S,\varphi ,T_c)\right) \,.
\label{eq:reducedhamilton}
\end{equation}

The invariance of the potential under the change of the total phase $\varphi_1+\varphi_2$ implies a conservation law, which in the $Z_\mu=0$ gauge allows us to work out the phase $\varphi_2$ in terms of the relative phase $\varphi$~\cite{Cline:2011mm}:
\begin{equation} 
\partial_z \varphi_2 = - \frac{h_1^2}{h_1^2+h_2^2} \partial_z \varphi \,.
\label{eq:h2phase}
\end{equation} 
The complex, spatially-varying top mass can now be constructed from the phase $\varphi_2(x)$ and the modulus $h_2(z)$:
\begin{equation} 
m_t(z) = \frac{y_t}{\sqrt{2}} h_2(z) e^{i\varphi_2(z)}.
\end{equation} 
In fact, one does not need to solve for the top phase, since only its derivative, given by Eq.~(\ref{eq:h2phase}), appears in the source term for the diffusion equations for chemical potentials:
\begin{equation}
S_t = \xi_w \left( 
    K_{8,t} (x^2_t\varphi_2')' -  K_{9,t}x_t^2 x_t^{2\prime}\varphi_2' 
          \right) \,.
\label{sourceterm}
\end{equation}
Here $\xi_w$ is the wall velocity, primes denote $\partial_{zT}$ and $K_{n,t}$ are dimensionless functions of $x_t \equiv |m_t|/T$ arising from phase-space averaging of certain kinematic variables defined in~\cite{Fromme:2006wx}. 
\begin{figure}
\begin{center}
\includegraphics[width=0.475\textwidth]{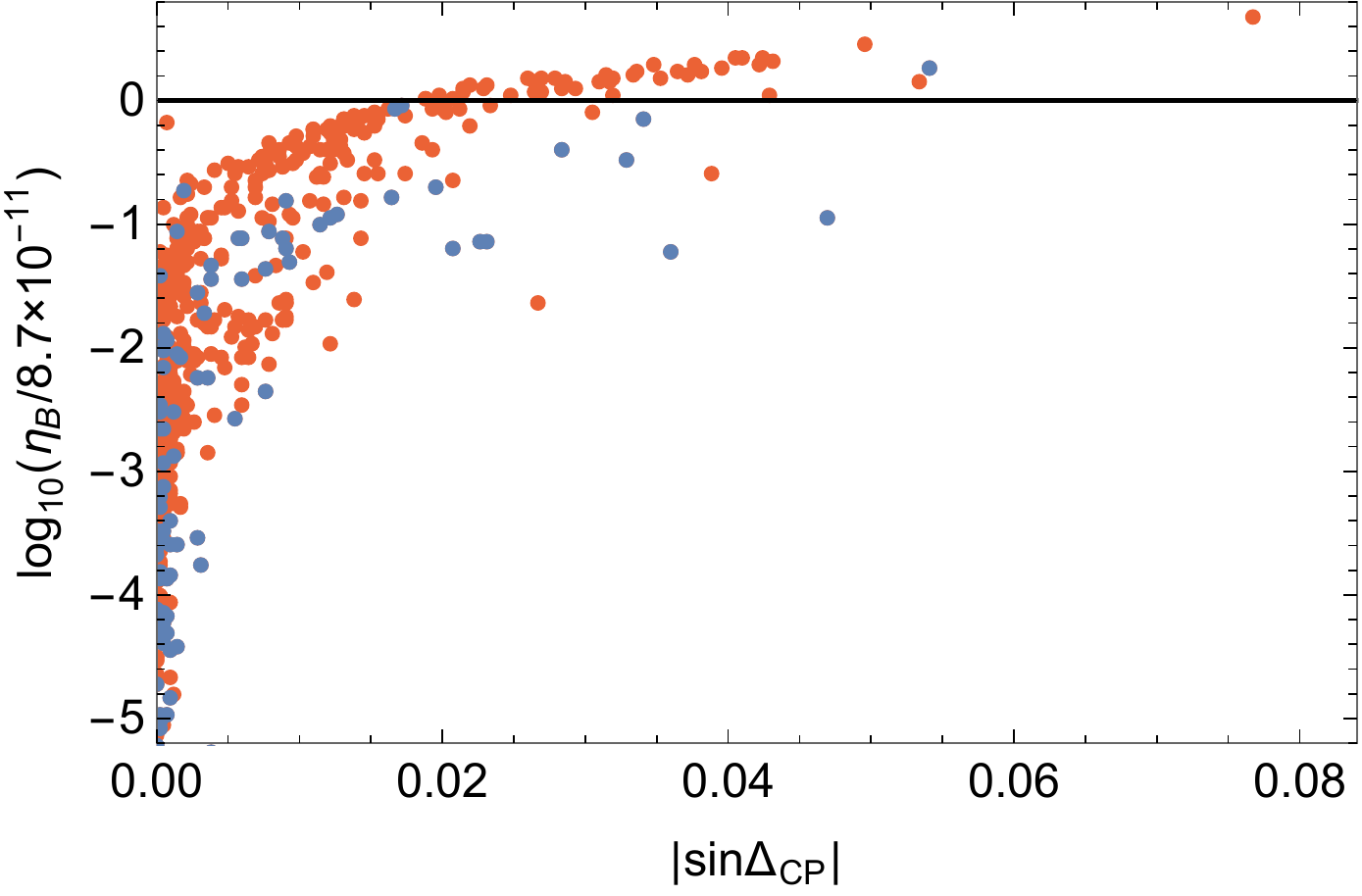} \hspace{1mm}
\includegraphics[width=0.470\textwidth]{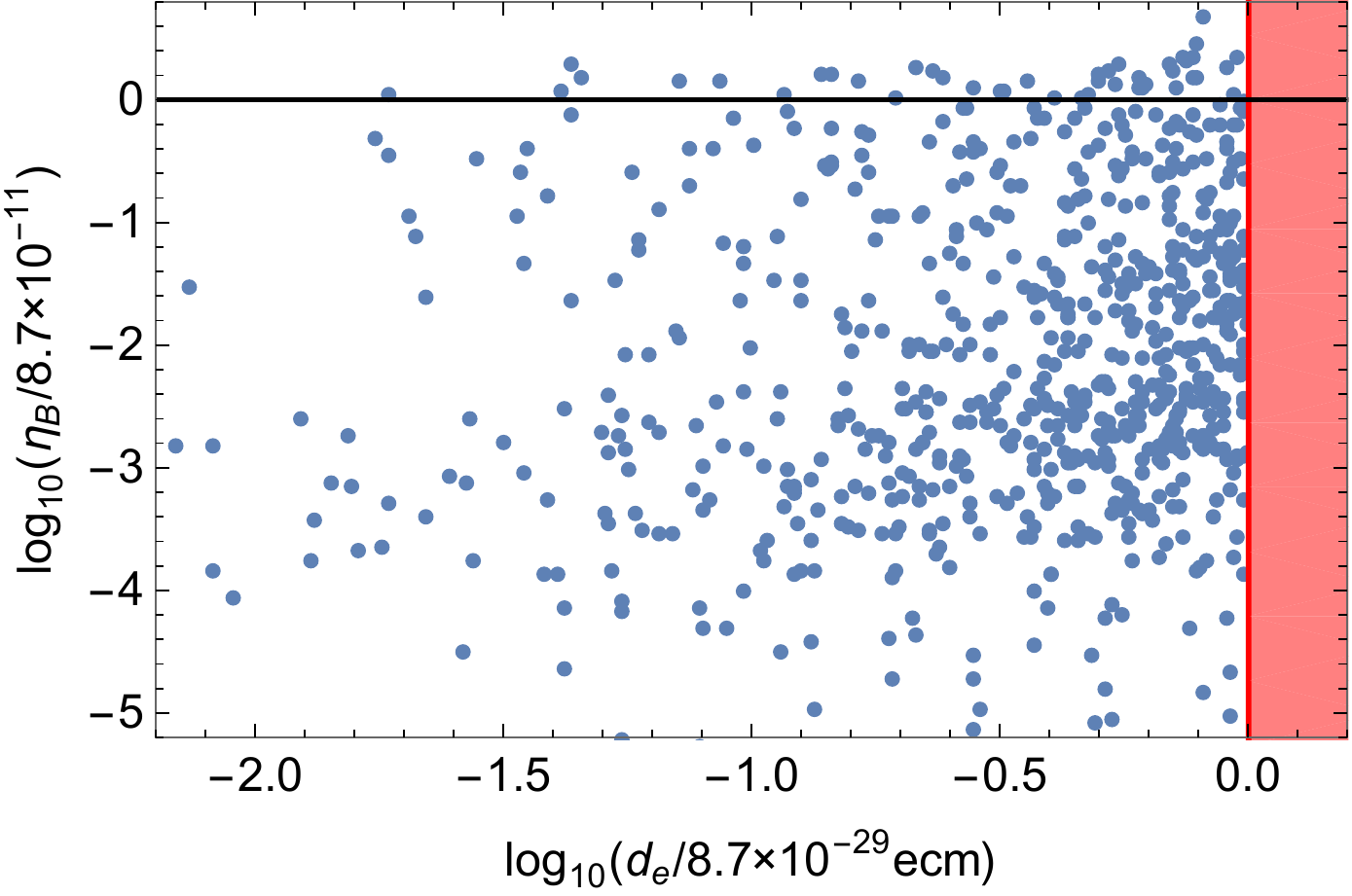}
\caption{Left: Shown is the correlation between the baryon-to-entropy ratio $\eta_B$ and the mixing matrix element $\sin\Delta_{\rm CP}$. Red dots correspond to models for which $T_n$ cannot be found (in the thin-wall approximation). Right: the correlation between $\eta_B$ and $d_e$. The red region is excluded by the eEDM limit and the black line shows the observed baryon-to-entropy ratio.}
\label{etaB}
\end{center}
\end{figure}

Given the source, one can calculate chemical potentials $\mu_j(z)$ for top, bottom, anti-top and Higgs by solving a set of transport equations defined in~\cite{Fromme:2006cm}. Finally the baryon-to-entropy ratio $\eta_B \equiv n_B/s$ is given by
\begin{equation} 
\eta_B  = \frac{405}{4\pi^2 \xi_w g_* T_c} \int_0^\infty {\rm d}z\, \Gamma_{\rm sph}(z) \mu_{B_L}(z) e^{-45 \Gamma_{\rm sph}(z) z/4\xi_w}.
\end{equation} 
We take $\xi_w = 0.1$ for the wall velocity and $g_*=106.75$ for the number of degrees of freedom in the plasma. The left-chiral baryon chemical potential is
\begin{equation} 
\mu_{B_L} = \frac{1}{2}(1+4K_{1,t})\mu_t + \frac{1}{2}(1+4K_{1,b})\mu_b - 2 K_{1,t_c} \mu_{t_c}.
\end{equation} 
For the sphaleron rate we use a formula interpolating between the symmetric and the broken phase~\cite{Cline:2011mm},
\begin{equation} 
\Gamma_{\mathrm{sph}}(z) = \min(10^{-6}T_c, 2.4T_c e^{-40 v(z)/T_c}),
\end{equation}
where $v(z)^2 = h_1(z)^2+h_2(z)^2$.

In the left panel of Fig.~\ref{etaB} we show how the baryon-to-entropy ratio relative to the observed value $\eta_B^{\rm obs} =  8.7\times 10^{-11}$~\cite{Ade:2015xua} correlates with the CP-violation-sensitive parameter $\sdcp$ defined in (\ref{eq:RN42}). Shown are only the points which survive the eEDM bound. As expected, the size of $\sdcp$ correlates with the size of $\eta_B$. This trend is similar to the correlation between $\sdcp$ and $d_e$ shown in Fig.~\ref{eEDM}. However, a large $\eta_B$ does not always imply a large $d_e$, as is clear from the right panel of Fig.~\ref{etaB}, where we show the correlation between $d_e$ and $\eta_B$, again for points that pass the EDM bound. Apparently, while both quantities are sensitive to the CP-violating parameters in the model, they can be sensitive to different linear combinations of them, so that large $\eta_B$ may be obtained simultaneously with a small enough $d_e$. 

\begin{figure}
\begin{center}
\includegraphics[width=0.98\textwidth]{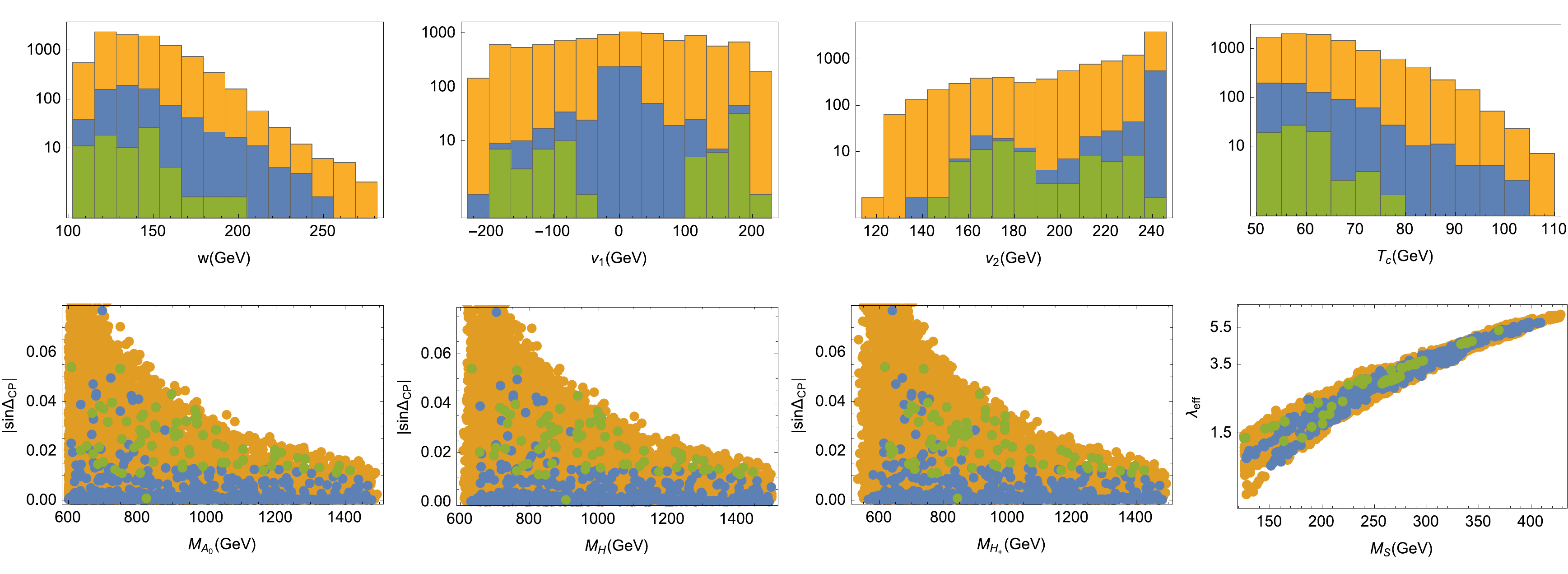}
\caption{Shown are the frequency distributions of the vevs of the scalar fields and the critical temperature as well as scatter plots for the masses of the new heavy scalar particles in our parameter scan. For details see the text.}
\label{histograms}
\end{center}
\end{figure}

Fig.~\ref{histograms} shows the distributions of various physical parameters in our parametric scan. Orange colour refers to models that pass all experimental cuts described in Secs. \ref{sec:experiments} and \ref{sec:DMlimits}, and give a strong EWPT, blue to models that in addition satisfy EDM constraint and green to models which also give large baryon-to-entropy ratio $\eta_B/\eta_B^{\rm obs} \in [0.5,2]$. These plots must be interpreted with care, since our scans were partly tuned by hand. Nevertheless, we see that none of the vevs can be very large and in particular both $v_1$ and $v_2$ need to be nonzero. Also the critical temperature is bounded from above: $T_c\lsim 100$ GeV. Finally for the models with large $\eta_B$, the new scalar masses are in general bound from above: $m_H, m_{A_0}, m_{H^\pm} \lsim 1.4$ TeV and $m_S \lsim 400$ GeV, which is encouraging from the point of view of experimental verifiability of the model.
%

%
\subsection{Bubble nucleation}
%

So far we have implicitly assumed that the bubble nucleation takes place at a temperature not too different from the critical temperature. This is typically the case in models where the first-order phase transition is effected by cubic corrections to potential from infrared modes, which leads to rather mild supercooling and small latent heat release. Here the situation is different, because the barrier between the degenerate minima is essentially due to a tree-level term. Thus a stronger supercooling and more latent heat release may be expected, or even a possibility of a formation of a metastable vacuum where the electroweak breaking never takes place. 
 
We study the nucleation problem in the thin-wall limit~\cite{Linde:1981zj}. The bubble nucleation rate is given by 
\begin{equation} 
\Gamma \sim T^4\left(\frac{S_3(T)}{2\pi T}\right)^{3/2}\exp\left( -\frac{S_3(T)}{T} \right),
\end{equation} 
where $S_3(T)$ is the three-dimensional action for an O(3)-symmetric bubble. In the thin-wall limit, it is given by
\begin{equation} 
S_3(T) = \frac{16\pi}{3} \frac{\sigma^3}{\Delta V(T)^2} \,,
\end{equation} 
where $\Delta V(T)$ is the potential energy difference between the electroweak-symmetric and electroweak-broken minima and $\sigma$ is the surface tension,
\begin{equation} 
\sigma = \int d\phi \sqrt{2V} \,,
\end{equation} 
integrated along the path from the symmetric to the broken minimum at temperature $T=T_c$. The bubble nucleation temperature $T_n$ is defined as the temperature at which creating at least one bubble per horizon volume is of order one. This condition can be written as
\begin{equation} 
\frac{S_3(T_n)}{T_n} = -\log\left(\frac{3}{4\pi} \Big(\frac{H(T_n)}{T_n}\Big)^4 \Big(\frac{2\pi T_n}{S_3(T_n)}\Big)^{3/2}\right) \,.
\label{eq:TnucEqn}
\end{equation} 

\begin{figure}
\begin{center}
\includegraphics[width=0.474\textwidth]{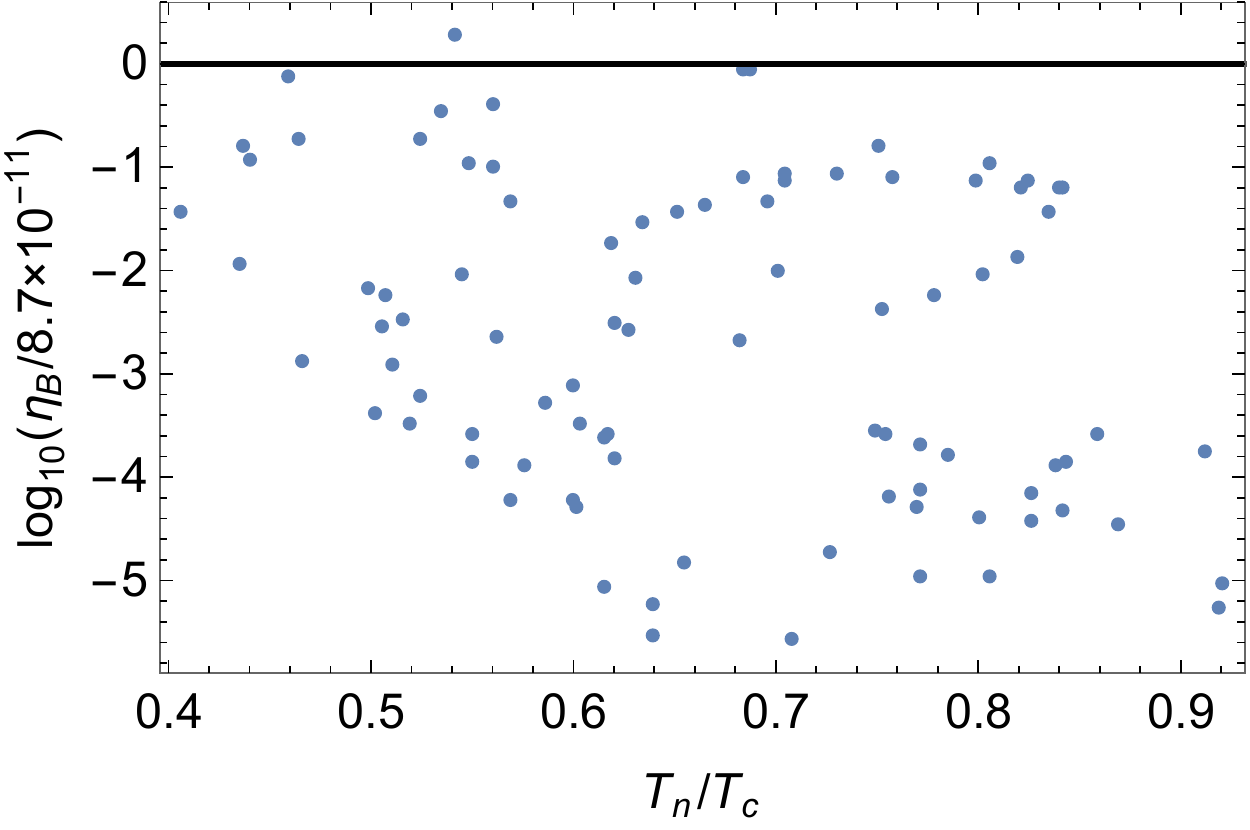} \hspace{1mm}
\includegraphics[width=0.476\textwidth]{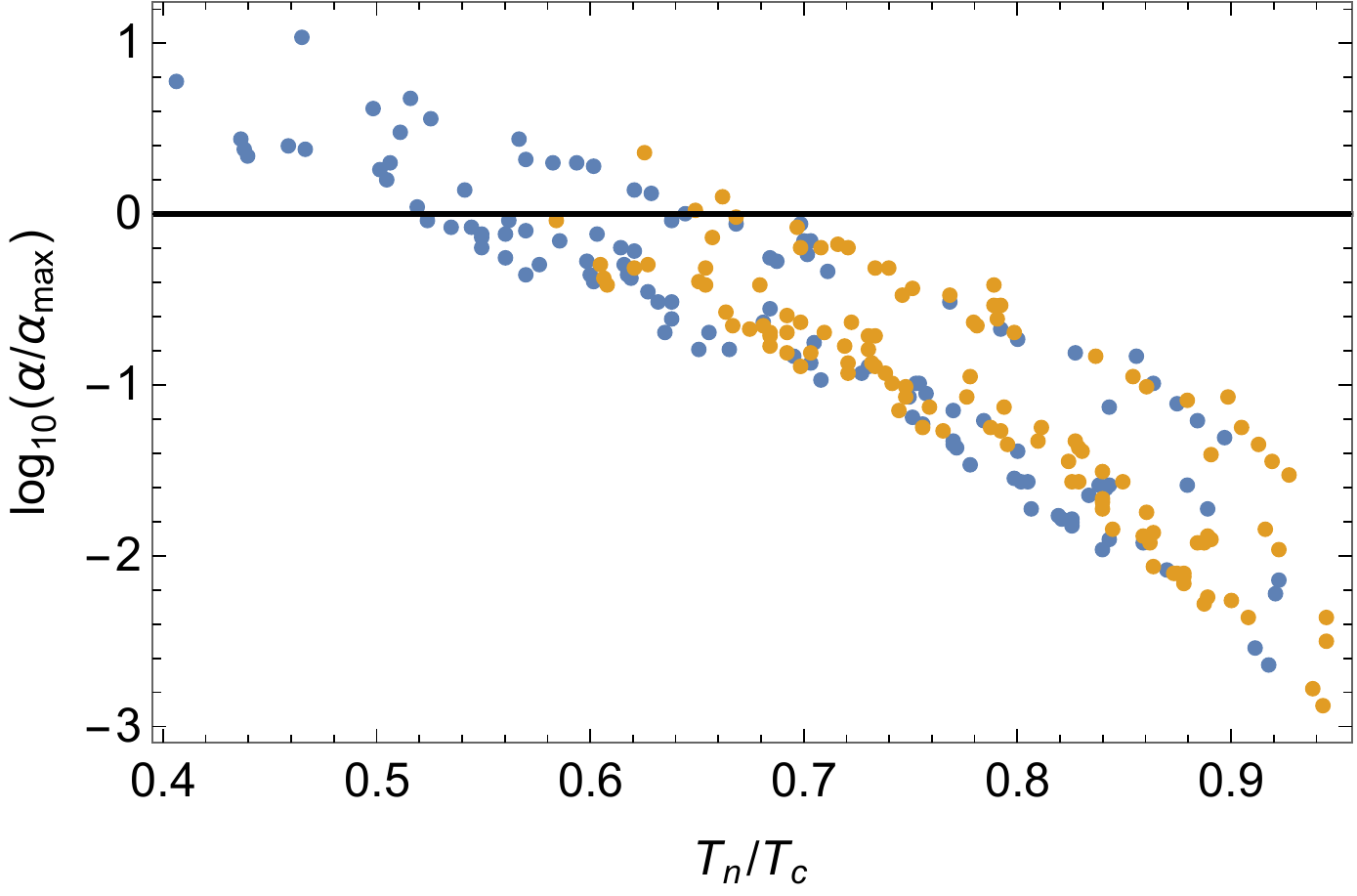}
\caption{Left: Shown is the baryon asymmetry $\eta_B$ as a function of the ratio of the nucleation and critical temperatures $T_n/T_c$.  Right: shown is the bound for finding a deflagration wall. Only models below the line $\alpha/\alpha_{\rm max}=1$ are allowed. For yellow points the nucleation temperatures were rescaled by $\kappa = 0.7$. See text for details. We used $\alpha_{\rm max}=0.38$, corresponding to $\xi_w\approx0.1$.} 
\label{fig:TnTc_etaB}
\end{center}
\end{figure}

We show the results of the nucleation-temperature calculation for our data set on the left panel of Fig.~\ref{fig:TnTc_etaB}. Obviously, a large number of points displayed in Fig.~\ref{etaB} are missing in Fig.~\ref{fig:TnTc_etaB}. The reason is that for these models, indicated by red dots in Fig.~\ref{etaB}, no solution to Eq.~(\ref{eq:TnucEqn}) was found. In these cases, in the thin-wall approximation, the fields were trapped in the false vacuum. Moreover, of the surviving models only four give large $\eta_B$. 

The situation is actually more dire than this: another implicit assumption in our baryon asymmetry calculation is that the transition walls are subsonic deflagrations, which is required for efficient diffusion of particle asymmetries across the bubble wall. However, with a large latent heat release the walls tend to be supersonic detonations instead. A full microscopic analysis of wall dynamics including a computation of the wall friction is beyond the scope of this paper. However, a deflagration wall must necessarily satisfy a condition~\cite{Espinosa:2010hh}
\begin{equation}
\alpha \equiv \frac{ \Delta V(T_n)}{\rho(T_n)}  <  \frac{1}{3} (1-\xi_w)^{-13/10} \equiv \alpha_{\rm max}\,,
\label{eq:deflagrationbound}
\end{equation}
where $\rho(T_n)$ is the radiation energy density in the symmetric phase. 
We show this condition in the right panel of Fig.~\ref{fig:TnTc_etaB} (blue dots) for the set of models shown in the left panel. As expected, for the models with the lowest nucleation temperatures, deflagrations are not possible. This applies in particular to all four surviving models with a large baryon-to-entropy ratio.

There are two issues that ameliorate the situation. First, the validity of the thin-wall limit actually requires a small latent heat and/or a large surface tension, which is often not the case here. When not applicable, thin-wall limit tends to overestimate the action $S_3(T)$, and hence underestimate the nucleation rate and eventually $T_n$. Accurate calculation of the nucleation rate is quite complicated in the full model, however, and we do not pursue it here. Instead, we compare the nucleation temperatures found in the thin-wall limit and in the full calculation in the simpler, singlet extension of the SM, studied for example in~\cite{Cline:2012hg,Cline:2013gha}. In practice we minimize the action
\begin{equation}
S_3(T) = 4\pi \int r^2 {\rm d}r \left(\frac{1}{2}\left(\frac{{\rm d}h}{{\rm d}r}\right)^2 + \frac{1}{2}\left(\frac{{\rm d}S}{{\rm d}r}\right)^2 + V_{\rm SSM}(h,S,T) \right) \,,
\label{eq:SSMS3}
\end{equation}
where $h$ is the SM-Higgs field, and $V_{\rm SSM}(h,S,T)$ is the singlet-model potential.

\begin{figure}
\begin{center}
\includegraphics[width=0.6\textwidth]{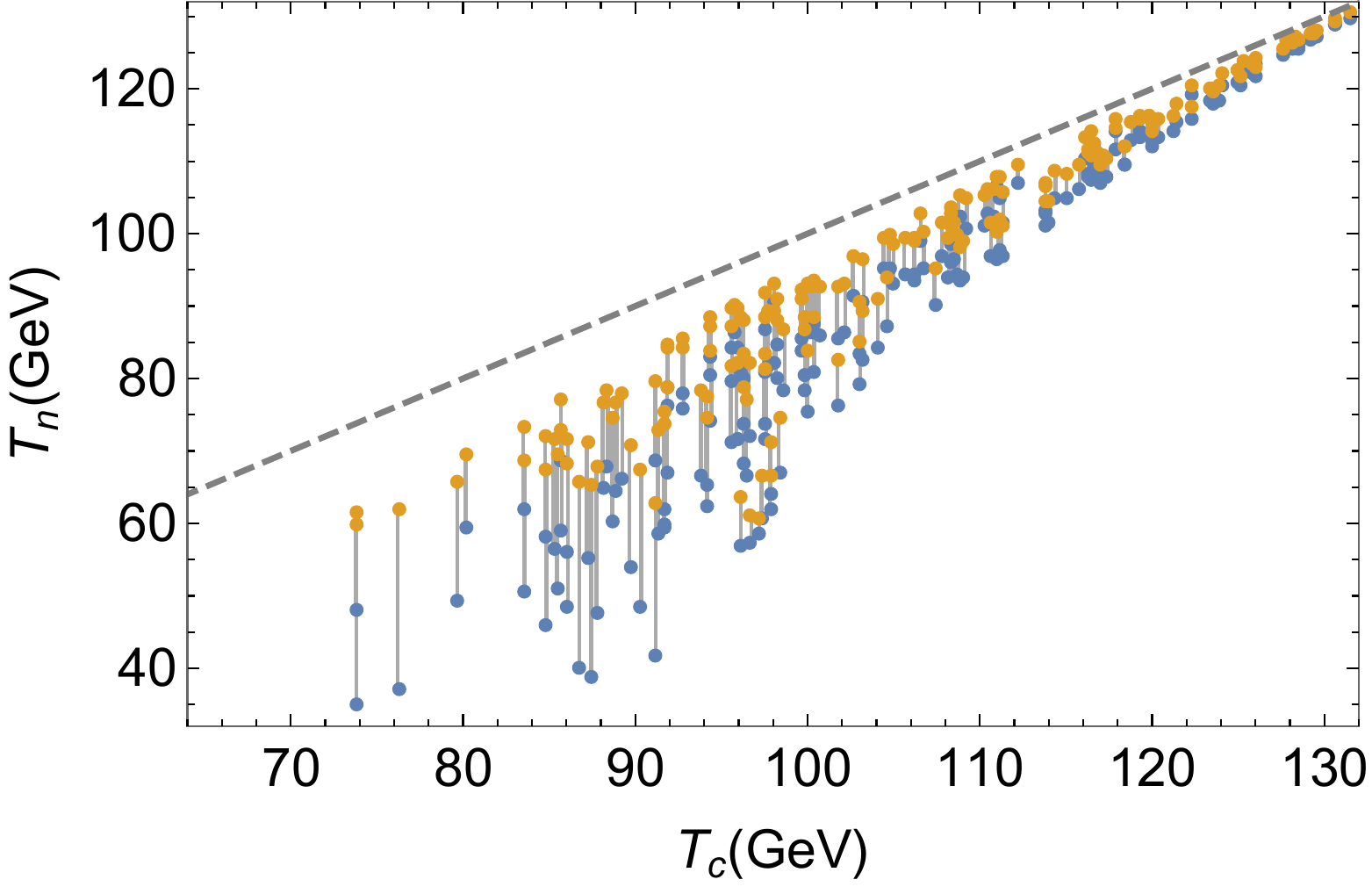} \hspace{0.1cm}
\caption{Shown are the nucleation temperatures in a scan over the parameters of the singlet extension of the SM. Blue dots correspond to the thin-wall temperature and the yellow dots to a full calculation. Gray lines connect pairs corresponding to the same physical parameters. The dotted line corresponds to $T_n=T_c$.} 
\label{fig:TcTn}
\end{center}
\end{figure}
The results of this analysis are shown in Fig.~\ref{fig:TcTn}. The blue dots show the nucleation temperature in the thin-wall approximation and the yellow dots the same quantity found from the solving minimizing the action \eqref{eq:SSMS3}\,.  As expected, the thin-wall approximation underestimates nucleation temperatures significantly. We find that the true nucleation temperature $T_n$ and the thin-wall value $T_n^{\rm tw}$ are related by $T_n = T_c - \kappa \,(T_c-T_n^{\rm tw})$, where the coefficient $\kappa$ to some extent depends on $M_S$ and $\lambda_S$, but is less than $0.7$. We anticipated this result in the deflagration limit shown in the right panel of Fig.~\ref{fig:TnTc_etaB}, where the yellow dots were found by redefining all thin-wall nucleation temperatures by the above equation with $\kappa=0.7$. We believe that this scaling conservatively represents the effect of going beyond thin-wall approximation in the full 2HDM and singlet model, and hence shows that most parameter sets may in fact be deflagrations.

\begin{figure}
\begin{center}
\includegraphics[width=0.478\textwidth]{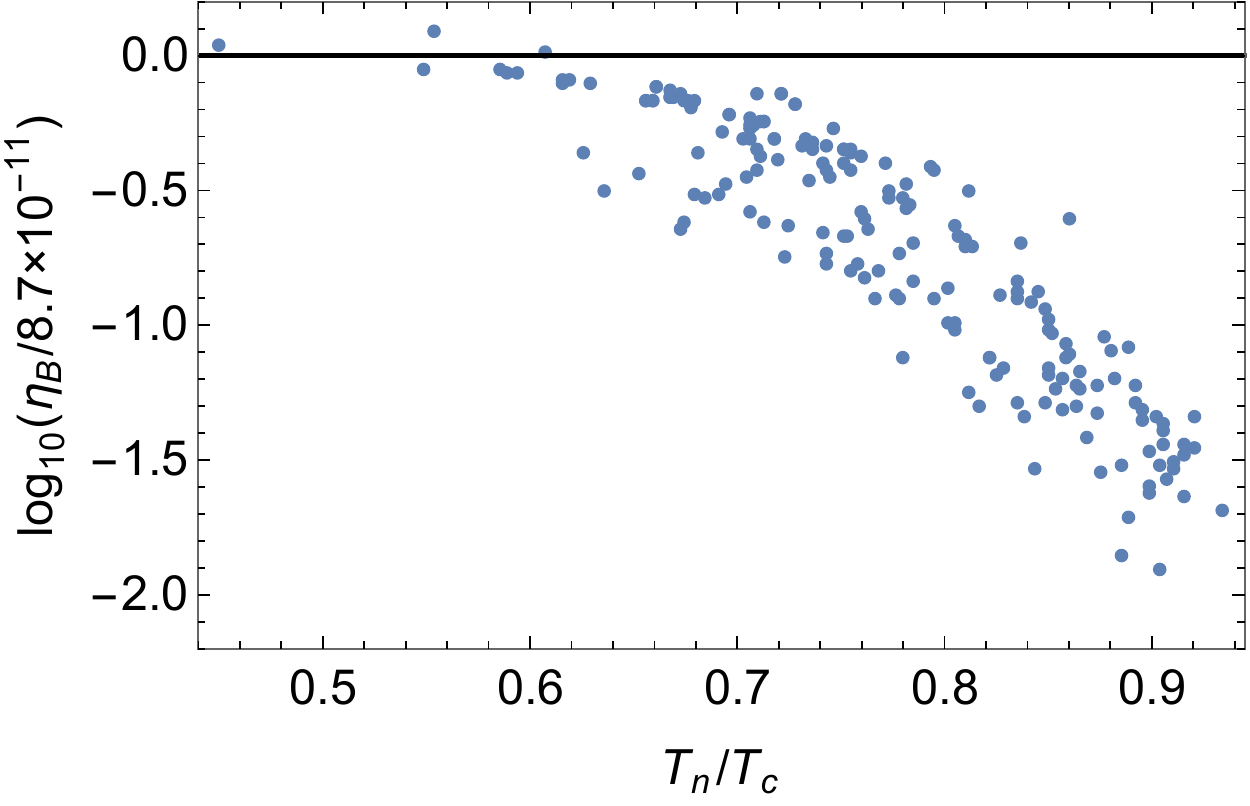} \hspace{1mm}
\includegraphics[width=0.47\textwidth]{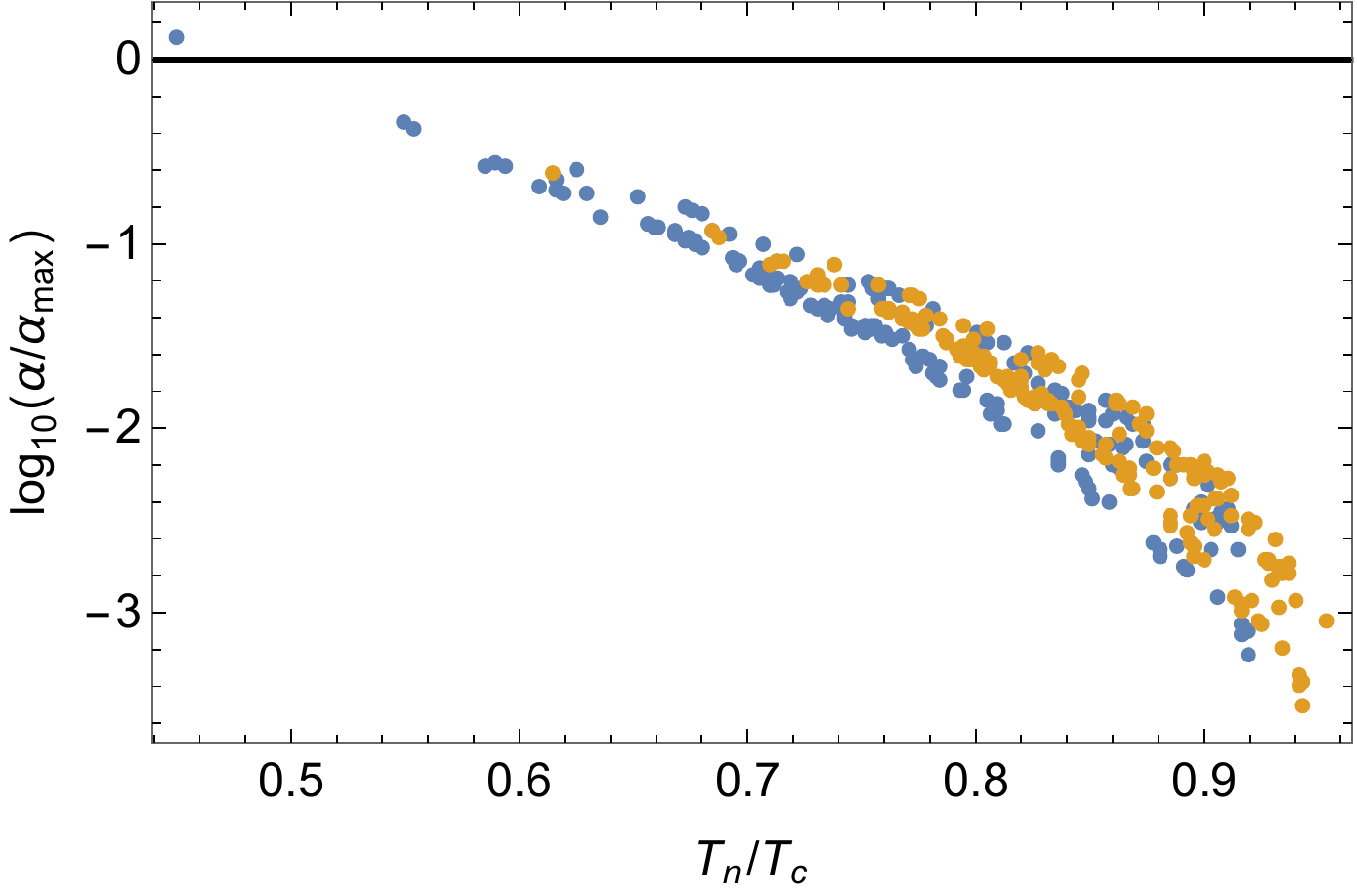}
\caption{Left: Shown is the baryon asymmetry $\eta_B$ as a function of $T_n/T_c$ for the new scan with $T_c>80$GeV (here $T_n$ is the thin-wall nucleation temperature). Right: the deflagration bound for the set displayed in the left plot as a function of $T_n/T_c$.  We again used $\alpha_{\rm max}=0.38$.} 
\label{fig:TcTnEtaB2}
\end{center}
\end{figure}

We can also tune our search to prefer models with a higher critical temperatures. Indeed, as is clearly seen from Fig.~\ref{fig:TcTn}, the nucleation temperature approaches the critical temperature when $T_c$ gets higher in the singlet extension of the SM and this feature persists also in the full 2HDSM. Hence, we made a new parametric scan, where we accept only models with $T_c>80$~GeV. The result of this scan is seen in Fig.~\ref{fig:TcTnEtaB2}, where the left panel again shows the baryon-to-entropy ratio and in the right panel the deflagration bound. We now found more points with large asymmetry and, in particular after one rescales the thin-wall nucleation temperatures as explained above, these points are now well below the deflagration bound Eq.~(\ref{eq:deflagrationbound}).

%

%
\section{Conclusions and outlook} \label{sec:conclusions}
%

We have studied the viability of a two-Higgs-doublet and inert-singlet model for EWBG and for DM, taking into account also all existing observational and collider constraints. Our model is based on the maximal GL(2,$\mathbbm{C}$) reparemetrization symmetry. This implies a universal Yukawa-alignment scheme, where both Higgs fields couple similarly to all fermions and there are no FCNCs. Exploiting the GL(2,$\mathbbm{C}$) symmetry, the the model can, in a particular basis, be written with a type-I Yukawa sector combined with the most general CP-violating potential. 

Following~\cite{Ivanov:2006yq,Ivanov:2007de,Ivanov:2008er} we implemented a novel way to construct potentials with a tree-level stability and to study the symmetry breaking patterns at finite temperatures. This construction was based on the Lorentz symmetry induced by the reprametrization symmetry on bilinears formed from Higgs doublets. These techniques are applicable to all 2HDM models, and they proved extremely useful when performing large-scale parametric scans over the multidimensional phase space of the model.

Dark matter and the strength of the electroweak phase transition in the model follow a similar pattern to the pure singlet extension: in accordance with~\cite{Cline:2012hg}, we find that strong two-step phase transitions are easily found, but they are consistent only with a subleading DM. Likewise, we find that experimental and observational constraints are fairly easy to avoid, with the outstanding exception of the electron-EDM bound, which strongly constrains the CP-violating parameters on the model. EDM constraints are particularly important because creating a large baryon asymmetry during the electroweak phase transition requires large CP-violating parameters; we found that the electron EDM indeed strongly constrains the phase space consistent with baryogenesis. Yet the bounds are not as strong here as in the pure 2HDM case~\cite{Cline:2011mm}, and we found a number of models consistent with all requirements.

Finally, we observed that two-step transitions may suffer from an unexpected problem of providing a {\em too strong} phase transition. We found that fields may get trapped in the metastable minimum, and transition walls may not be subsonic as required by a successful baryogenesis scenario. However, our analysis in the full model was restricted to the thin-wall approximation. We  then studied the bubble nucleation in full generality in the case of a pure singlet extension of the SM. While the generic problem of too strong transitions remained, we found that thin-wall limit tends to overestimate the strength of the transition. Based on these results we argued that, when corrected for the thin-wall bias, the walls may well remain subsonic in the full 2HDSM. In a revised scan concentrating to models with a large critical temperature, we found many models potentially consistent with all available constraints.

While our results are not a definite proof, they provide a strong indication of a success of baryognesis in the 2HDM and an inert singlet model. Settling the issue beyond any doubt would require two significant improvements. First is a detailed analysis of the bubble wall dynamics, including a microscopic computation of the friction on the wall. Second is going beyond the $T_c$-bounce solution, when solving the scalar field profiles over the bubble wall to compute the top quark mass profile and eventually the baryogenesis source. These are both very interesting topics that deserve to be studied in detail in the future.

%
\section*{Acknowledgemets}
%

We thank Jim Cline for useful comments and discussions and collaboration at the early stages of this work. KK would like to thank Stephan Huber, Mikko Laine and Jose M. No for useful discussions during the 2016 MIAPP programme ``Why is there more Matter than Antimatter in the Universe?''. This work was financially supported by the Academy of Finland projects 278722 and 267842. The CP$^3$-Origins centre is partially funded by the Danish National Research Foundation, grant number DNRF90. VV is supported by the Magnus Ehrnrooth foundation. TA  acknowledges partial funding from a Villum foundation grant.

%
\appendix
%

%
\section{Diagonalization of mass matrices}
\label{massappendix}
%

In this appendix we show in detail the diagonalization of the scalar mass matrices. Following \cite{Degee:2009vp}, we write
\begin{equation}
r^\mu = \phi_x \Sigma_{xy}^\mu\phi_y \,,
\end{equation}
where
\begin{equation}
\phi = \left( H_{1I}^+, H_{2I}^+, H_{1R}^+,  H_{2R}^+, H_{1I}^0, H_{2I}^0, H_{1R}^0, H_{2R}^0 \right) \,,
\end{equation}
where the subscripts $R$ and $I$ again denote the real and imaginary parts, respectively, and the matrices $\Sigma^\mu$ are block diagonal matrices consisting of $4\times 4$ elements
\begin{equation}
\Sigma^0 = 1_4 \,,\, 
\Sigma^1 = \left(\begin{array}{cc} \sigma_1 & 0 \\ 0 & \sigma_1 \end{array}\right) \,, \,
\Sigma^2 = \left(\begin{array}{cc} 0 & -i \sigma_2 \\ i \sigma_2 & 0\end{array}\right) \,,\, 
\Sigma^3 = \left(\begin{array}{cc} \sigma_3 & 0 \\ 0 & \sigma_3\end{array}\right) \,.
\end{equation}
Now the $(a,b)$ element of the mass matrix is
\begin{equation}
\left.\frac{\partial^2 V}{\partial \phi_a \partial \phi_b}\right|_{\phi = \langle \phi \rangle} = - \Sigma_{ab}^\mu M_\mu + \Sigma_{ab}^\mu \lambda_{\mu\nu} \langle r^\nu \rangle + 2 \Sigma_{ax}^\mu \langle \phi_x \rangle \lambda_{\mu\nu} \langle \phi_y \rangle \Sigma_{yb}^\nu \,,
\end{equation}
and the minimum conditions are
\begin{equation}
\left.\frac{\partial V}{\partial \phi_a}\right|_{\phi = \langle \phi \rangle} = \left( - \Sigma_{ax}^\mu M_\mu + \Sigma_{ax}^\mu \lambda_{\mu\nu} \langle r^\nu \rangle \right) \langle \phi_x \rangle = 0\;.
\end{equation}
Let us consider the neutral part of the mass matrix. The eigenstate corresponding to eigenvalue zero is $G_0 = N_{G} \Sigma^5 \langle \phi^0 \rangle$\,, where $\Sigma^5 = \Sigma^0 \Sigma^1 \Sigma^2 \Sigma^3$ and $N_{G}$ is the normalization factor. Now
\begin{equation}
G_0 = \cos\beta \cos\theta \, H_{1R}^0 + \sin\beta \, H_{1I}^0 - \cos\beta\sin\theta \, H_{2R}^0 \,,
\end{equation}
where 
\begin{equation}
\beta = \arctan{\frac{v_2}{v_1}}\,.
\end{equation}
Hence we can diagonalize the neutral mass matrix by
\begin{equation}
R_N = \left(
\begin{array}{cccc}
1 & 0 & 0 & 0 \\
0 & c _y c_z &  -c_y s_z  & s_y \\
0 & s_x s_y c_z+c_x s_z & c_x c_z-s_x s_y s_z & -s_x c_y \\
0 & s_x s_z-c_x s_y c_z & c_x s_y s_z+s_x c_z & c_x c_y
\end{array} \right) \left(
\begin{array}{cccc}
c_{\beta}c_{\theta} & s_{\beta} & -c_{\beta}s_{\theta} & 0 \\
-s_{\beta}c_{\theta} & c_{\beta} & s_{\beta}s_{\theta} & 0 \\
s_{\theta} & 0 & c_{\theta} & 0 \\
0 & 0 & 0 & 1
\end{array}
\right)\,,
\label{eq:neutraldiagonalizingmatrix}
\end{equation}
where we have used the short-hand notations $s_{\alpha}\equiv \sin\alpha$, $c_{\alpha}\equiv\cos\alpha$, and the neutral mass eigenstates are
\begin{equation}
\left(\begin{array}{c} G_0 \\ A_0 \\ H_0 \\ h_0\end{array}\right) = R_N \left(\begin{array}{c} H_{1I}^0 \\  H_{2I}^0 \\  H_{1R}^0 \\ H_{2R}^0\end{array}\right)\;.
\end{equation}

On the charged sector there are two eigenstates with eigenvalue zero: $G_{1} = N_{G_1} \Sigma^5 \langle \phi^0 \rangle  $ and $G_{2} = N_{G_2} \langle \phi^0 \rangle$\;. Now
\begin{equation}
\begin{split}
&G_{1} = \cos\beta \sin\theta \, H_{1R}^+ + \cos\beta \cos\theta \, H_{1I}^+ + \sin\beta \, H_{2I}^+ \,, \\
&G_{2} = \cos\beta \cos\theta \, H_{1R}^+ - \cos\beta \sin\theta \, H_{1I}^+ + \sin\beta \, H_{2R}^+ \,.
\end{split}
\end{equation}
Hence charged Goldstone bosons are $G_\pm = \cos\beta e^{\pm i\theta} H_1^\pm + \sin\beta H_2^\pm$\,. The charged mass matrix in the basis $\{H_1^+, H_2^+\}$ can be diagonalized by
\begin{equation}
R_C = \left(\begin{array}{cc}c_\beta e^{i \theta} & s_\beta \\-s_\beta e^{i \theta} & c_\beta\end{array}\right)\;,
\end{equation}
and charged mass eigenstates are
\begin{equation}
\left(\begin{array}{c} G_+ \\ H_+ \end{array}\right) = R_C \left(\begin{array}{c} H_{1}^+ \\  H_{2}^+ \end{array}\right)\;.
\end{equation}
%

%
\section{1-loop beta functions for scalar couplings}
\label{RGEs}
%

For completeness we show the 1-loop beta functions,
\begin{equation}
    \label{eq:}
    \beta_{\lambda_i}\equiv\frac{\mathrm{d} \lambda_i}{\mathrm{d}\ln \mu},    
\end{equation}
we used to compute the perturbativity limits on the model parameters:

\begin{align}
\begin{split}
16\pi^2\beta_{\lambda_1}
	    = & \;\;24\lambda_1^2 + 2\lambda_3^2 + 2\lambda_3\lambda_4 + \lambda_4^2
	        + 4|\lambda_5|^2 + 12|\lambda_6|^2\\
          & +\sfrac{1}{2}\lambda_{S1}^2 
           - 3\left(3g_{\mathrm{L}}^2+g_Y^2\right)\lambda_1 +\sfrac{3}{8}\left(3g_{\mathrm{L}}^4
  		   + 2g_{\mathrm{L}}^2g_Y^2+g_Y^4\right),
\end{split} \\
\begin{split}	    
16\pi^2\beta_{\lambda_2}
        = &\;\; 24\lambda_2^2 + 2\lambda_3^2 + 2\lambda_3\lambda_4 + \lambda_4^2
           + 4|\lambda_5|^2 + 12|\lambda_7|^2 + \sfrac{1}{2}\lambda_{S2}^2  \\
	      &-3\left(3g_{\mathrm{L}}^2 + g_Y^2\right)\lambda_2 
	       + \sfrac{3}{8}\left(3g_{\mathrm{L}}^4 + 2g_{\mathrm{L}}^2g_Y^2 + g_Y^4\right)
	       +12y_t^2\lambda_2-6y_t^4,
\end{split}\\
\begin{split}	    
16\pi^2\beta_{\lambda_3}
        = &\;\;(\lambda_1+\lambda_2)(12\lambda_3+4\lambda_4)+4\lambda_3^2+2\lambda_4^2
		  +8|\lambda_5|^2+4|\lambda_6|^2\\
	      &+16\mathrm{Re}\left(\lambda_6^*\lambda_7\right)+4|\lambda_7|^2+\lambda_{S1}\lambda_{S2}\\
	      &-3\left(3g_{\mathrm{L}}^2+g_Y^2\right)\lambda_3
		 +\sfrac{3}{4}\left(3g_{\mathrm{L}}^4-2g_{\mathrm{L}}^2g_Y^2+g_Y^4\right)+6y_t^2\lambda_3,
\end{split}\\
\begin{split}
16\pi^2\beta_{\lambda_4}
        = &\;\; 4(\lambda_1+\lambda_2)\lambda_4+8\lambda_3\lambda_4+4\lambda_4^2
		+ 32|\lambda_5|^2+10\left(|\lambda_6|^2+|\lambda_7|^2\right)\\
	    & +4\mathrm{Re}\left(\lambda_6^*\lambda_7\right)+|\lambda_{S12}|^2
		-3\left(3g_{\mathrm{L}}^2+g_Y^2\right)\lambda_4+3g_{\mathrm{L}}^2g_Y^2
		+6y_t^2\lambda_4,
\end{split}\\
\begin{split}	   
16\pi^2\beta_{\lambda_5}
        = &\;\; (4\lambda_1+4\lambda_2+8\lambda_3+12\lambda_4)\lambda_5+5\lambda_6^2
	    + 2\lambda_6\lambda_7+5\lambda_7^2\\
	    & +\sfrac{1}{2}\lambda_{S12}^2
	    -3\left(3g_{\mathrm{L}}^2+g_Y^2\right)\lambda_5+6y_t^2\lambda_5,
\end{split}\\
\begin{split}
16\pi^2\beta_{\lambda_6}
        =&\;\; (24\lambda_1+6\lambda_3+8\lambda_4)\lambda_6
		+(6\lambda_3+4\lambda_4)\lambda_7+(20\lambda_6^*+4\lambda_7^*)\lambda_5\\
	    &+\lambda_{S1}\lambda_{S12}-3\left(3g_{\mathrm{L}}^2+g_Y^2\right)\lambda_6+3y_t^2\lambda_6,
\end{split}\\
\begin{split}
16\pi^2\beta_{\lambda_7}
        =&\;\; (6\lambda_3+4\lambda_4)\lambda_6
		+(24\lambda_2+6\lambda_3+8\lambda_4)\lambda_7+(4\lambda_6^*+20\lambda_7^*)\lambda_5\\
	    &+\lambda_{S2}\lambda_{S12}-3\left(3g_{\mathrm{L}}^2+g_Y^2\right)\lambda_7+9y_t^2\lambda_7,
\end{split}\\
\begin{split}
16\pi^2\beta_{\lambda_S}
        =&\;\; 18\lambda_S^2+2\lambda_{S1}^2+2\lambda_{S2}^2+4|\lambda_{S12}|^2,
\end{split}\\
\begin{split}
16\pi^2\beta_{\lambda_{S1}}
        =&\;\; (12\lambda_1+6\lambda_S)\lambda_{S1}+(4\lambda_3+2\lambda_4)\lambda_{S2}
		+4\lambda_{S1}^2+12\mathrm{Re}\left(\lambda_6^*\lambda_{S12}\right)\\
	    &+4|\lambda_{S12}|^2-\sfrac{3}{2}\left(3g_{\mathrm{L}}^2+g_Y^2\right)\lambda_{S1},
\end{split}\\
\begin{split}
16\pi^2\beta_{\lambda_{S2}}
        =&\;\; (4\lambda_3+2\lambda_4)\lambda_{S1}+(12\lambda_2
		+6\lambda_S)\lambda_{S2}+4\lambda_{S2}^2+12\mathrm{Re}\left(\lambda_7^*\lambda_{S12}\right)\\
	    &+4|\lambda_{S12}|^2-\sfrac{3}{2}\left(3g_{\mathrm{L}}^2
	    +g_Y^2\right)\lambda_{S2}+6y_t^2\lambda_{S2},
\end{split}\\
\begin{split}
16\pi^2\beta_{\lambda_{S12}}
        =&\;\; (2\lambda_3+4\lambda_4+6\lambda_S+4\lambda_{S1}
		+4\lambda_{S2})\lambda_{S12}+12\lambda_5\lambda_{S12}^*\\
	    &+6\lambda_6\lambda_{S1}+6\lambda_7\lambda_{S2}-\sfrac{3}{2}\left(3g_{\mathrm{L}}^2
		+g_Y^2\right)\lambda_{S12}+3y_t^2\lambda_{S12}.
\end{split}
\end{align}


%
\section{Singlet scalar annihilation cross sections} \label{cross-sections}
%

Here we give compact expressions for singlet pair annihilation rates used in the computation of the DM abundances.  In the mass eigenbasis, the relevant Lagrangian may be written as
\begin{equation}
\begin{split}
\mathcal{L} =& \sum_{j,k} \left(c_0 S^2 H_j H_k + \sum_i c_{ijk} H_i H_j H_k \right) + \sum_i c_i S^2 H_i + \sum_{i,f} \left( y_{if} H_i \overline{f}_{\mathrm{R}} f_{\mathrm{L}} + \text{h.c.} \right) \\
&+ \sum_i \left(g_{Wi} W^+ W^- H_i + g_{Zi} Z Z H_i\right) \,,
\end{split}
\end{equation}
where $H = \left( A_0, H_0, h_0, H^+ \right)$\,. Now the singlet scalar annihilation cross section to scalars is given by:
\begin{equation}
\sigma(SS\to H_jH_k) = \frac{1}{16\pi s^2 v_s} \left( A_{jk}^2 I_{0,j,k} + 2 A_{jk} B_{jk} I_{1,j,k} + B_{jk}^2 I_{2,j,k} \right)\,.
\end{equation}
The rate into fermion--antifermion pairs is given by:
\begin{eqnarray}
\sigma(SS\to \overline{f}f) &=& \frac{I_{0,f,f}}{4\pi s^2 v_s} \sum_{i,j} 
\frac{c_i c_j}{(s-m_i^2)(s-m_j^2)} \times
\nonumber \\ \hskip2.5truecm
&&\times \left( \left( y_{ifR} y_{jfR} - y_{ifI} y_{jfI} \right) \frac{s}{2} - 2y_{ifR} y_{jfR} m_f^2 \right)\,,
\end{eqnarray}
and the annihilation to vector bosons is:
\begin{equation}
\sigma(SS\to VV) = \frac{I_{0,V,V}}{4\pi s^2 v_s} \sum_{i,j}  \frac{c_i c_j g_{Vi} g_{Vj}}{(s-m_i^2)(s-m_j^2)}\left( 3+\frac{s(s-4M_V^2)}{4M_V^2} \right) \left(1-\frac{\delta_{V,Z}}{2}\right)  \,,
\end{equation}
where
\begin{equation}
\begin{split}
&I_{0,j,k} = \sqrt{s^2+(m_j^2-m_k^2)^2 - 2 s (m_j^2+m_k^2)} \,, \\
&I_{1,j,k} = \frac{4 I_{0,j,k}}{m_j^2+m_k^2-s} \,, \\
&I_{2,j,k} = \frac{2 I_{0,j,k}}{m_j^2 m_k^2 - M_S^2 (2m_j^2 + 2m_k^2 - s) + \frac{M_s^2}{s}(m_j^2-m_k^2)^2} + \frac{8 I_{0,j,k}}{(m_j^2+m_k^2-s)^2} \,,
\end{split}
\end{equation}
and
\begin{equation}
\begin{split}
&A_{jk} = 2(1+\delta_{jk})c_0 + 2\sum_i (1+\delta_{ij}+\delta_{jk}+\delta_{ki}+2\delta_{ij}\delta_{jk}) \frac{c_i c_{ijk}}{s-m_i^2} \,,\\
&B_{jk} = 4 c_j c_k \,.
\end{split}
\end{equation}
%

%
\bibliography{twoDS.bib}
%

\end{document}